\documentclass[acmsmall,authorversion]{acmart} 

\usepackage{comment}
\usepackage{subfig} 
\usepackage{multirow} 
\usepackage{pbox} 
\usepackage{tabularx}
\usepackage{soul,xcolor}
\usepackage{graphics}
\AtBeginDocument{%
  \providecommand\BibTeX{{%
    \normalfont B\kern-0.5em{\scshape i\kern-0.25em b}\kern-0.8em\TeX}}}

\setcopyright{rightsretained} 
\acmJournal{PACMHCI}
\acmYear{2022} \acmVolume{6} \acmNumber{CSCW2} \acmArticle{505} \acmMonth{11} \acmPrice{}\acmDOI{10.1145/3555618}

\received{January 2022}
\received[revised]{April 2022}
\received[accepted]{August 2022}

\begin{document}
\setstcolor{red}
\title[Characterization, Detection, and Personalization of Religiously Intolerant Arabic Videos]{Deradicalizing YouTube: Characterization, Detection, and Personalization of Religiously Intolerant Arabic Videos}


\author{Nuha Albadi}
\orcid{0000-0003-1273-8371}
 \affiliation{%
 \institution{Taibah University}
 \department{Department of Computer Science}
 \country{Saudi Arabia}}
 \affiliation{%
 \institution{University of Colorado Boulder}
 \department{Department of Computer Science}
 \country{USA}}
 \email{nuha.albadi@colorado.edu}

\author{Maram Kurdi}
 \affiliation{%
 \institution{Taif University}
 \department{Department of Computer Science}
 \country{Saudi Arabia}}
 \affiliation{%
 \institution{University of Colorado Boulder}
 \department{Department of Computer Science}
 \country{USA}}
 \email{maram.kurdi@colorado.edu}
 
\author{Shivakant Mishra}
\affiliation{%
 \institution{University of Colorado Boulder}
 \department{Department of Computer Science}
 \country{USA}}
\email{mishras@colorado.edu}




\begin{abstract}
Growing evidence suggests that YouTube's recommendation algorithm plays a role in online radicalization via surfacing extreme content. Radical Islamist groups, in particular, have been profiting from the global appeal of YouTube to disseminate hate and jihadist propaganda. In this quantitative, data-driven study, we investigate the prevalence of religiously intolerant Arabic YouTube videos, the tendency of the platform to recommend such videos, and how these recommendations are affected by demographics and watch history. Based on our deep learning classifier developed to detect hateful videos and a large-scale dataset of over 350K videos, we find that Arabic videos targeting religious minorities are particularly prevalent in search results (30\%) and first-level recommendations (21\%), and that 15\% of overall captured recommendations point to hateful videos. Our personalized audit experiments suggest that gender and religious identity can substantially affect the extent of exposure to hateful content. Our results contribute vital insights into the phenomenon of online  radicalization and facilitate curbing online harmful content.


\end{abstract}
%



\maketitle


\section{Introduction}
YouTube, the largest video hosting website \cite{YouTubelargest}, has been under intense scrutiny from both the media \cite{NyTimesYouTubeRadical,NyTimesYouTubetheGreat,WashingtonYouTubeRadical,TheWallStreetJournal} and the academic community \cite{Radicalization_Pathways,conway2008jihadi,rochert2020homogeneity,o2015down,murthy2021evaluating} for allegedly serving as a radicalizing instrument driving people down rabbit holes of radical and extreme content. As with any other content personalization system, YouTube's recommendations and search results serve {\it personalized} content tailored to individual users based on online users' behavior and personal data \cite{YouTubeMetrics,covington2016deep}. A major concern often associated with content personalization systems is creating what is commonly referred to as the ``filter bubble'' effect, in which users get served content that reinforces their beliefs and social identity, while content from opposing viewpoints and perspectives gets filtered \cite{pariser2011filter}. These radicalization and filter bubble accusations against YouTube have motivated numerous CSCW/HCI researchers to \textit{audit} YouTube's recommendation algorithm to investigate its tendency to surface and steer users toward problematic content \cite{hussein2020measuring,Kostantinos2022justAflue,Radicalization_Pathways,papadamou2020disturbed}. Some of these audit studies accounted for personalization by creating multiple user accounts with different watch histories and personal data \cite{Kostantinos2022justAflue,hussein2020measuring}, while others assessed the platform's recommendation system for a generic (non-personalized) user account \cite{Radicalization_Pathways,papadamou2020disturbed,rochert2020homogeneity}. The focus of these studies was on misinformation 
\cite{hussein2020measuring}, conspiracy theories \cite{Kostantinos2022justAflue}, and radicalization within a Western context, e.g., the radical right \cite{Radicalization_Pathways,rochert2020homogeneity}. Evidence shows that radical Islamist groups have been benefiting from the unparalleled growth in YouTube audiences to proliferate jihadist propaganda and other radical ideologies \cite{klausen2012youtube,conway2008jihadi,murthy2021evaluating}. Yet, auditing YouTube for Islamist radical content remains a largely unexplored research area.


To address this gap, this paper presents a large-scale, quantitative study to investigate the prevalence of Arabic YouTube videos promoting hate against religious minorities, the tendency of YouTube's recommendation algorithm to surface and recommend such content, and how these recommendations are affected by users' watch history and demographics. Our study focuses on six religious groups: Muslims, Christians, Jews, atheists, Sunni (the most prominent Islamic denomination), and Shia (the second-largest Islamic denomination). To perform a thorough assessment of YouTube's algorithm, we conduct two types of analysis; 1) non-personalized analysis, where we assess YouTube's recommendation algorithm for a logged-out user scenario; 2) personalized audits, in which we audit YouTube for several logged-in user accounts with different personal attributes. 

For our non-personalized analysis, our paper answers the following research questions: 
\begin{itemize}
  \item[] \textbf{RQ1} What is the extent of religiously intolerant Arabic videos? Which religious groups are targeted the most?
  \item[] \textbf{RQ2} How often does YouTube's recommendation algorithm recommend hateful videos? Does it recommend hateful videos to non-hateful ones?
  \end{itemize}
  
To account for personalization in our assessment, our paper answers the following research question:
  
  \begin{itemize}
  \item[] \textbf{RQ3} What is the effect of personalization based on (a) religious ideology (moderate vs. radical), (b) Islamic denomination (Sunni vs. Shia), and (c) gender (female vs. male) on the volume of hateful content presented to users in search results and recommendations?
\end{itemize}


We used YouTube's API to collect data for our first two research questions concerning our non-personalized analysis. It has been shown that videos returned by the API were similar to those displayed to a logged-out user browsing YouTube \cite{Kostantinos2022justAflue}. Thus, using the API, we collected a seed list of 3,000 Arabic YouTube videos discussing various religious groups. Using crowdworkers, we acquired annotations for these videos to identify the ones promoting religious hate along with the targeted religious group(s). For each video in our seed list, we collected its top four recommendations going five levels deep, which resulted in a total of over 350K unique recommended videos. Leveraging our annotated dataset, we developed a deep learning classifier that can effectively detect hateful YouTube videos with an accuracy of 0.76. We applied our classifier to the 350K recommended videos to estimate the prevalence of religiously intolerant Arabic videos. Finally, using traced recommendation relationships between these videos, we created a directed graph that resembled YouTube's recommendation graph to measure the tendency of YouTube's recommendation algorithm to suggest hateful content. 

While our first two research questions explore religiously intolerant Arabic content on YouTube for a generic (non-personalized) user account, our third research question seeks to understand the {\it effect of personalization} on the volume of hateful videos presented to users with different demographics and watch history. In particular, we investigate the impact religious ideology (moderate vs. radical), Islamic denomination (Sunni vs. Shia), and gender (female vs. male) on the level of exposure to hateful videos in search results and recommendations. To this end, we carefully crafted eight different user profiles, each with a distinctive set of personal attributes. To establish the Islamic denomination and religious ideology attributes, we slowly built a watch history for each of the eight profiles in a controlled environment over the course of nine weeks (more on the methodology in Section \ref{personalization_audits}). We then conducted two systematic audit experiments, \textit{Search} and \textit{Recommendation} audits, to explore whether personalization based on religious ideology, Islamic denomination, and gender has any significant effect on the levels of exposure to hateful content. 

The following presents a summary of our key findings:
\begin{itemize}
    \item Our deep learning classifier can effectively detect hateful Arabic videos with an F\textsubscript{1} score of 0.69, a promising first attempt result. We make our ground truth dataset public\footnote{\url{https://osf.io/cf9w8/?view_only=aa81f43ff28c4faaa7514ccccc6a386c}} for use by the research community. 
  \item We find that videos promoting religious hatred in the Arabic language are prevalent on search results (29.53\%) and first-level recommendations (20.80\%), and they are mainly targeting Shia and atheists (\textbf{RQ1}). 
  \item We find that recommendations pointing to hateful videos represent about 15\% of overall recommendations. Further, about 12\% of videos recommended to non-hateful videos are hateful, while about 31\% of videos recommended to hateful videos are also hateful (\textbf{RQ2}).
   \item We observe a religious ideology effect on recommendations for Shia profiles, in which \textit{radical} Shia profiles were recommended 29\% more hateful videos than \textit{moderate} Shia profiles (\textbf{RQ3a}). 
   \item We find that moderate \textit{Sunni} profiles were recommended 21\% more hateful videos than moderate \textit{Shia} profiles, indicating an Islamic denomination effect (\textbf{RQ3b}). 
   \item We observe a gender effect on videos recommended to Sunni profiles, in which \textit{male} Sunni profiles were recommended significantly more hateful videos (16\% increase) than \textit{female} Sunni profiles (\textbf{RQ3c}). 
   \item We find that personalization in general increases the risk of getting recommended hateful videos by 46\% compared to recommendations obtained from YouTube API, which are not personalized (\textbf{RQ3}). 
\end{itemize}

\section{Background and Literature Review}
This section provides background information on practices around religion, culture, and norms within Arab countries to help the reader better understand some of the concepts and findings presented in this paper. Additionally, to situate our research within the CSCW literature, we review scholarly works on online hate and extremism and algorithmic auditing.

\subsection{Arabs and Religion}

The Middle East and North Africa (MENA) region consists of twenty Arab countries. Muslims make up 93\% of the region's total population, followed by Christians (3.7\%), Jews (1.6\%), and atheists (0.6\%) \cite{pew2015religious}. The two major denominations of Islam are Sunni (87-90\%) and Shia (10-13\%) \cite{Pewmapping}. While Muslims are predominantly of the Sunni denomination, Shia comprises the majority of Muslims in Bahrain (70\%), Iraq (65\%), and Lebanon (55\%) \cite{Pewmapping}. 

To better comprehend the study methodology and results, it is essential to understand the central religious role of clerics within an Islamic context. An Islamic cleric, also known as imam, mufti, and sheik, refers to a religious leader who oversees worship, interprets religious texts, and passes fatwa, i.e., a ruling under Islamic (Shari'ah) law \cite{mouline2014clerics}. An Islamic cleric more closely resembles in their sacred role and status a Jewish rabbi than a Christian priest \cite{pipes2017path}. Although Muslim clerics don't serve as mediators between God and people, many Muslims feel the need to follow clerics' rules and teachings even on the most private matters \cite{pipes2017path}. In general, Islamic clerics are regarded as well-respected community leaders, with some of them having a TV show and/or YouTube channel where they educate, teach, and advise people on matters of faith and everyday life. However, while many Muslim clerics adopt a moderate interpretation of Islam, some hold a more radical view of Islam centered around the ideology of militant jihad \cite{nielsen2017deadly}.

Middle Eastern societies have long suffered from civil wars and domestic tensions that are partly caused by conflicting religious beliefs \cite{mihaylov2017arab,PewSocial}. The Sunni-Shia divide dates back to the seventh century \cite{marshall2010islamic}, and it is believed to have ignited several wars in the region \cite{gonzalez2013sunni}. International conflicts between Arabs (predominantly Sunni) against Israelis (mostly Jewish) and Iranians (mainly Shia) have religious aspects in addition to political, economic, and ethnic ones \cite{cardinali2013sunni,fraser2015arab}. Apostasy, i.e., renouncing one's religious belief, is considered a major crime deserving capital punishment in five MENA countries \cite{REUTERSathiests}. This religious tension, among other factors, contributes to the region's civil unrest and sectarian violence. 

However, numerous academic institutions and government agencies have devoted efforts to promoting tolerance and acceptance of stigmatized religious minorities in the region in recent years. For instance, in 2017, Saudi Arabia established Etidal \footnote{\url{https://etidal.org/en/home/}}, a global center for combating extremist ideology by providing counter-narratives that promote coexistence, moderation, and acceptance values. Arabic research raising awareness and mitigating online hate speech has also grown in number in recent years \cite{Albadi2019investigating,mulki2019hsab,haddad2019t,chowdhury2019arhnet,farha2020multitask,omar2020comparative,albadi2019hateful,faris2020hate, alshalan2020deep,al2021detection, magdy2016failedrevolutions}. Our work extends these efforts by investigating the prevalence of religiously intolerant Arabic YouTube videos and building automated detection models to hinder their reach. 

\subsection{Online Radicalization and Hate Speech}

 Online hate speech has been extensively studied in a Western context across multiple protected characteristics such as race \cite{kwok2013locate,burnap2016us,chandrasekharan2017you,mathew2019thou}, gender \cite{badjatiya2017deep,frenda2019online,samory2021call}, religion \cite{magdy2016isisisnotislam,zannettou2020quantitative,mathew2019thou,ashrafyoutube}, sexual orientation \cite{burnap2016us,lingiardi2020mapping,mathew2019thou}, and disability \cite{burnap2016us,lingiardi2020mapping}. Some studies focused on identifying targets of hate in online social networks \cite{lingiardi2020mapping,silva2016analyzing,mondal2017measurement}, while others measured the growth and reach of hate speech over time \cite{mathew2020hate,mathew2019spread}. Multiple machine learning techniques and algorithms have been explored such as Naive Bayes \cite{kwok2013locate} support vector machines \cite{badjatiya2017deep,magdy2016isisisnotislam}, regressions \cite{badjatiya2017deep}, decision trees \cite{kaati2015detecting,badjatiya2017deep}, and neural networks \cite{badjatiya2017deep,Albadi2019investigating,zhang2018detecting}. Multiple text representation techniques have been used ranging from character n-grams \cite{waseem2016hateful} to word and paragraph embeddings \cite{badjatiya2017deep,djuric2015comment}. Our work extends this line of research by investigating the discriminative power of visual features extracted from video thumbnails in distinguishing online hateful videos.

 Although hate speech and radicalization have been studied in multiple social network sites such as Twitter, Reddit, Gab.com (a loosely moderated social network), and Whisper (an anonymous social network), it has been rarely studied on YouTube (exceptions include \cite{mariconti2019you,ottoni2018analyzing,mathew2019thou,agarwal2014focused}). The researchers in \cite{ottoni2018analyzing} observed that right-wing YouTube channels contain a high volume of hateful content against Muslims and the LGBTQ community. In \cite{mariconti2019you}, the researchers developed a detection model to identify videos on YouTube that are being attacked by third-party coordinated hate raids. \citet{mathew2019thou} investigated the use of counter-speech in comments to tackle hateful YouTube videos. YouTube comments have been studied for other related issues such as toxicity \cite{obadimu2019identifying}, harassment \cite{aggarwal2014mining}, cyberbullying \cite{mouheb2018detection}, and moderation in general \cite{kurdi2021think}.

There exists limited CSCW work that investigates radicalization within an Islamist context, distinguishable exceptions include \cite{kursuncu2019modeling,Albadi2019investigating,albadi2019hateful}. \citet{kursuncu2019modeling} modeled Islamist extremist communications on Twitter along three dimensions: religion, ideology, and hate. The measurement study conducted by \citet{Albadi2019investigating} concluded that 42\% of Arabic tweets discussing other religions  incited hatred toward religious minorities in the region. In their follow-up study \cite{albadi2019hateful}, they found that Twitter bots (i.e., automated accounts) were responsible for 11\% of those hateful tweets. Our work builds on and extends these CSCW scholarships by providing important insights into Islamist radicalization and religious hate speech on YouTube.

\subsection{Algorithmic Auditing}
It is well-known that all major social media, including YouTube, feed personalized content to their users based on their online activities and collected personal data. According to YouTube, these personal data include watch and search history, age, gender, geographic location, the content of subscribed channels, time of the day, and other metrics related to the quality of the video (e.g., whether or not other users watched the entirety of the video) \cite{YouTubeMetrics,covington2016deep}. YouTube has also stated that more than 70\% of the total viewing time on their platform is the result of a recommendation-driven viewing \cite{YouTubeTotalWatchTime}. 

 Content personalization systems have been said to have a filter bubble effect, which refers to the intellectual and ideological isolation created by content personalization systems that trap an individual in a bubble of like-minded content, isolating them from other perspectives and viewpoints \cite{pariser2011filter}. These unintended risks of content personalization systems have intensified the need for algorithmic auditing, which refers to the "process of investigating the functionality and impact of decision-making algorithms" \cite{mittelstadt2016automation}. YouTube's filter bubble and radicalization claims have motivated the CSCW/HCI community to assess the platform's tendency to surface and steer users toward far-right radical content \cite{Radicalization_Pathways}, misinformation and conspiracy theories \cite{hussein2020measuring,Kostantinos2022justAflue}, and disturbing videos targeting kids \cite{papadamou2020disturbed}. Some audit studies accounted for personalization by creating Google accounts with different watch history \cite{Kostantinos2022justAflue,hussein2020measuring}, whereas others performed random walks without being logged into a Google account \cite{Radicalization_Pathways,papadamou2020disturbed,rochert2020homogeneity}.
 
On the other hand, there is a limited body of work that refutes these algorithmic radicalizations and filter bubble claims, particularly for online news consumption on YouTube and other search engines \cite{bakshy2015exposure,hosseinmardi2021examining,haim2018burst,robertson2018auditing}. For example, \citet{bakshy2015exposure} studied Facebook news consumption patterns and found that exposure to ideologically opposing views is more dependent on the individual choice rather than the ranking algorithm. In a recent study \cite{hosseinmardi2021examining}, the authors examined radical news consumption on YouTube and found that most people who consume far-right videos arrive at such videos from search results, the home page, or an external website rather than following YouTube recommendation chains. In an empirical study \cite{haim2018burst}, the authors explored the effect of personalization on the homogeneity of Google News and found no evidence to support the filter bubble phenomenon. Our study extends these audit experiments to include an Islamist radicalization context by measuring the effect of personalization on the degree of exposure to religiously intolerant videos in the Arabic language.  


\section{Data and Prior Analysis}

\subsection{Data Collection}

Our study is mainly concerned with YouTube videos promoting religious hate in the Arabic language, and thus we focus on the most common religious beliefs among Arabs. These are Islam, Christianity, Judaism, atheism, and the two main denominations of Islam: Sunni and Shia. In September 2019, we used YouTube API v3 \footnote{\url{https://developers.google.com/youtube/v3/}} to collect data for this part of the study. YouTube API doesn't account for personalization \cite{Kostantinos2022justAflue}, i.e., videos returned by the API are not affected by watch history or any other personal data, but rather are based on content relevance to the search query and other quality and user engagement metrics. 

\begin{table}[t]
\centering
\begin{minipage}[t]{0.48\linewidth}\centering
 \caption{Number of collected search result videos per religious group.}
 \label{search_results}
 \begin{tabular}{lc}
    \toprule
    Religious group & Number of videos \\
    \midrule
    Christians & 1,172 \\ 
    Muslims  & 1,003 \\ 
    Shia  & 850 \\ 
    Jews  & 782  \\ 
    Sunni  & 739  \\ 
    Atheists  & 646  \\ 
    \midrule
    All religious groups & 5,192\\
  \bottomrule
\end{tabular}
\end{minipage}\hfill%
\begin{minipage}[t]{0.48\linewidth}\centering
 \caption{Number of unique videos within each recommendation level.}
 \label{recommendation_results}
  \begin{tabular}{lc}
    \toprule
    Level \# & Number of unique videos \\
    \midrule
    Level 1 & 8,069 \\ 
    Level 2  & 23,319 \\ 
    Level 3  & 60,466 \\ 
    Level 4  & 137,789  \\ 
    Level 5  & 286,190  \\ 
    \midrule
    All levels  & 351,262  \\ 
  \bottomrule
\end{tabular}
\end{minipage}
\end{table}

To assess of the prevalence of religiously intolerant Arabic videos in YouTube's search results, we queried YouTube API using 208 impartial Arabic keywords\footnote{\url{https://osf.io/cf9w8/?view_only=aa81f43ff28c4faaa7514ccccc6a386c}} that refer to each of the aforementioned religious beliefs/groups. We manually compiled these keywords and made sure they don't include any religious slurs, hate terms, or insults that could bias search results. For example, for Muslims, the used keywords translate to: a Muslim [singular masculine], a Muslim [singular feminine], Muslims [plural masculine], Muslims [plural feminine], Muslims [dual masculine], Muslims [dual feminine], Islam, the Islam, the religion of Islam, Islamic religion. The reason for having what seems like a large number of keywords is that Arabic nouns can be either singular, dual, or plural; each can be either masculine or feminine. Additionally, Arabic nouns can have different spellings, and thus we accounted for all possible variations in spellings.

YouTube API allows for selecting a sorting method by which search result videos in the API response can be sorted. This can be based on either date, relevance (to the search query), rating, and view count. We selected relevance, which is also the default sorting method when accessing YouTube from a browser. For each keyword, we collected up to 50 most relevant Arabic videos. For each video, we collected its metadata (e.g., title, description, number of views, and number of likes), thumbnail, and up to 100 most recent comments. Table 
\ref{search_results} summarizes the number of videos collected per set of keywords. In total, we collected 5,192 unique videos; of these 1,172 were collected for Christian-related keywords, 1,003 for Muslim-related keywords, 850 for Shia-related keywords, 782 for Jew-related keywords, 739 for Sunni-related keywords, and 646 for atheist-related keywords. For each religious group, we randomly selected 500 videos to be annotated (see Section~\ref{sec:annotation}) as hateful or non-hateful religious videos, which resulted in creating a \textit{ground truth dataset} of 3,000 videos.

Next, to measure the prevalence of religiously intolerant Arabic videos in YouTube recommendations, we used a cascaded approach where we first used YouTube API to collect the top four videos recommended for each video in our ground truth dataset. This resulted in 8,069 unique {\it level 1 recommended videos}. We then repeated this process of getting top four recommended videos for each video in level 1 to create {\it level 2 recommended videos}, which turned out to have 23,319 unique videos. We repeated this process further to create {\it level 3}, {\it level 4} and {\it level 5} recommended videos, consisting of 60,466, 137,789 and 286,190 unique videos respectively for a total of 351,262 unique recommended videos across all five levels (refer to Table \ref{recommendation_results}). This also resulted in capturing 929,596 recommendation links between videos.


\subsection{Data Annotation}
\label{sec:annotation}
To create a ground truth dataset of hateful and non-hateful religious videos, we used Appen \cite{Appen}, a crowdsourcing service that is known for having Arabic-speaking annotators \cite{mubarak2016demographic}, to get annotations for the 3,000 videos in our ground truth dataset. We provided annotators with the definition of religious hate speech as outlined in \cite{albadi2018they}, ``{\it a speech that is insulting, offensive, or hurtful and is intended to incite hate, discrimination, or violence against an individual or a group of people on the basis of religious beliefs or lack thereof.}'' We also provided annotators with examples of videos that should be classified as hateful (e.g., a video promoting a belief that disbelievers will burn in hell), non-hateful (e.g., a documentary about the Jewish culture), or unrelated to religion. Specifically, we asked annotators to watch enough of the video until they reach a judgment on whether to classify it as: a)\textit{ hateful}, if the purpose of the video seems to be an incitement of hatred, intolerance, or violence against one or more religious groups; b)\textit{ non-hateful,} if the video is related to religions, but its purpose is not a promotion of religious hatred/intolerance; or c) \textit{unrelated}, if the video is either deleted, unrelated to any religion, or in a non-Arabic language with no Arabic subtitles. If annotators decided that a video was hateful, they were asked a second question to specify one or more religious groups that the video was targeting.

Three different annotators judged each video. To ensure high-quality annotations, we first created a set of 190 test questions (refer to Appendix \ref{appendix}) consisting of evidently hateful, non-hateful, and unrelated videos from our seed videos that the first two authors reviewed, discussed, and agreed on their annotation. For annotators to qualify for the task, they had to pass an initial test consisting of five test questions with a minimum accuracy score of 80\%, i.e., they had to answer at least 4/5 test questions correctly. Each page of work contained five videos to be annotated, one of which was a hidden test question. Annotators needed to maintain the 80\% accuracy score throughout the task. Those who couldn't keep that accuracy were disqualified, and their annotations were excluded. To ensure that annotators actually watched part of the video before assigning annotations, we set a minimum time needed for annotators to finish a page of work. To specify this minimum time, we kept track of the time it took the first two authors to annotate each of the 190 test questions, and we found that it takes 1.2 minutes to annotate a video on average. To allow some flexibility, we set the minimum time for an annotator to complete a page of work consisting of five videos to five minutes. Annotators who finished a page of work in less than five minutes were disqualified, and their annotations were excluded. To make sure we only have Arabic-Speaking annotators, Google Translate was disabled, and the language for the task was set to Arabic. We paid each annotator 35 cents for completing a page of work. Given that completing a page of work was estimated to take 5-6 minutes, our average hourly pay ranged from \$3.5 to \$4.2, which is slightly above the average hourly pay reported for a similar platform, Amazon Mechanical Turk \cite{hara2018data}. We also note that our annotators were located within Arab countries, primarily Egypt and Algeria; our pay was 4x the minimum pay in these countries. Upon completing the task, annotators were offered to participate in a task satisfaction survey. A total of 21 annotators participated in the survey. Overall, the task was rated 4/5 based on pay, clarity of instructions, fairness of test questions, and ease of the job. The pay, in particular, was rated 4.1/5.

In total, the annotation process was carried out by 151 different annotators. The interquartile mean of the time it took annotators to review a video was 1.3 minutes. Appen provides an inter-annotator agreement score for each question, reflecting the level of agreement between annotators' answers weighted by their accuracy scores \cite{Appen_confidence}. The first annotation question that asked to specify whether a video was hateful or not had an average agreement score of 0.84, which is considered an almost perfect agreement. The second annotation question regarding identifying targeted religious groups had an average agreement score of 0.46, a moderate agreement which is expected for a question with seven options.

\textbf{Ethical Concerns.} Although our institution's IRB confirmed that our study doesn't require IRB review, we acknowledge the potential risks of exposing annotators to potentially hateful content. Thus, in the annotation task description, we provided a warning for annotators regarding possible hateful, violence, and/or radical content that might be included in the videos they are about to watch. Additionally, annotators were able to quit the task at any time and get paid for their work.


\subsection{Analysis of Ground Truth Dataset}
\label{groundtruth}

Here, we derive insights from our ground truth dataset, reporting the distribution of hateful and non-hateful videos, targeted religious groups, and relationships between hateful videos and hateful comments. 

To identify the distribution of hateful, non-hateful, and unrelated videos in our ground truth dataset, we considered the answer to the first annotation question with the highest confidence score. Confidence score is a score that Appen provides with each possible answer that reflects the level of agreement between annotators on that answer weighted by their accuracy scores. For example, if we have three annotators with different accuracy scores, judge a video with three possible answers (e.g., hateful, non-hateful, unrelated). Then, each answer would be accompanied by a confidence score that reflects Appen's confidence in each answer based on the number of annotators who selected that answer weighted by their accuracy scores \cite{Appen_confidence}. In our ground truth dataset, 29.53\% of the videos were found to be hateful, 52.97\% non-hateful, and 17.50\% unrelated. Examples of unrelated videos include videos discussing an acute condition as the word atheism in Arabic can also mean `acute', videos in the Persian language, a language different from the Arabic language but uses similar alphabets, and videos in the English language with no Arabic subtitles.

To identify religious groups most targeted by hateful videos, we considered answer(s) to the second annotation question, that asks annotators to identify targeted religious groups, with a confidence score of 0.3 or higher. Since we had three annotators, selecting a lower threshold would result in having videos classified as hateful with no targeted religious group(s) (i.e., cases where annotators agreed that a video was hateful, but each selected a different targeted religion). Note that answers to the second annotation question were only used for analysis in this section, i.e., they were not considered in our classifier or any later analysis. In analyzing religious groups targeted by hateful videos, we found that the most targeted religious group is the Shia, with about 34\% of hateful videos targeting them, closely followed by atheists (33\%). Christians ranked third with about 18\% of hateful videos targeting them, followed by Muslims (16\%). Jews (10.4\%) and Sunni (9.5\%) were among the least targeted religious groups. These percentages add up to more than 100\% as some videos were targeting more than one religious groups.

\begin{figure}[t]
\subfloat[Shia dataset]{\includegraphics[width=0.33\linewidth]{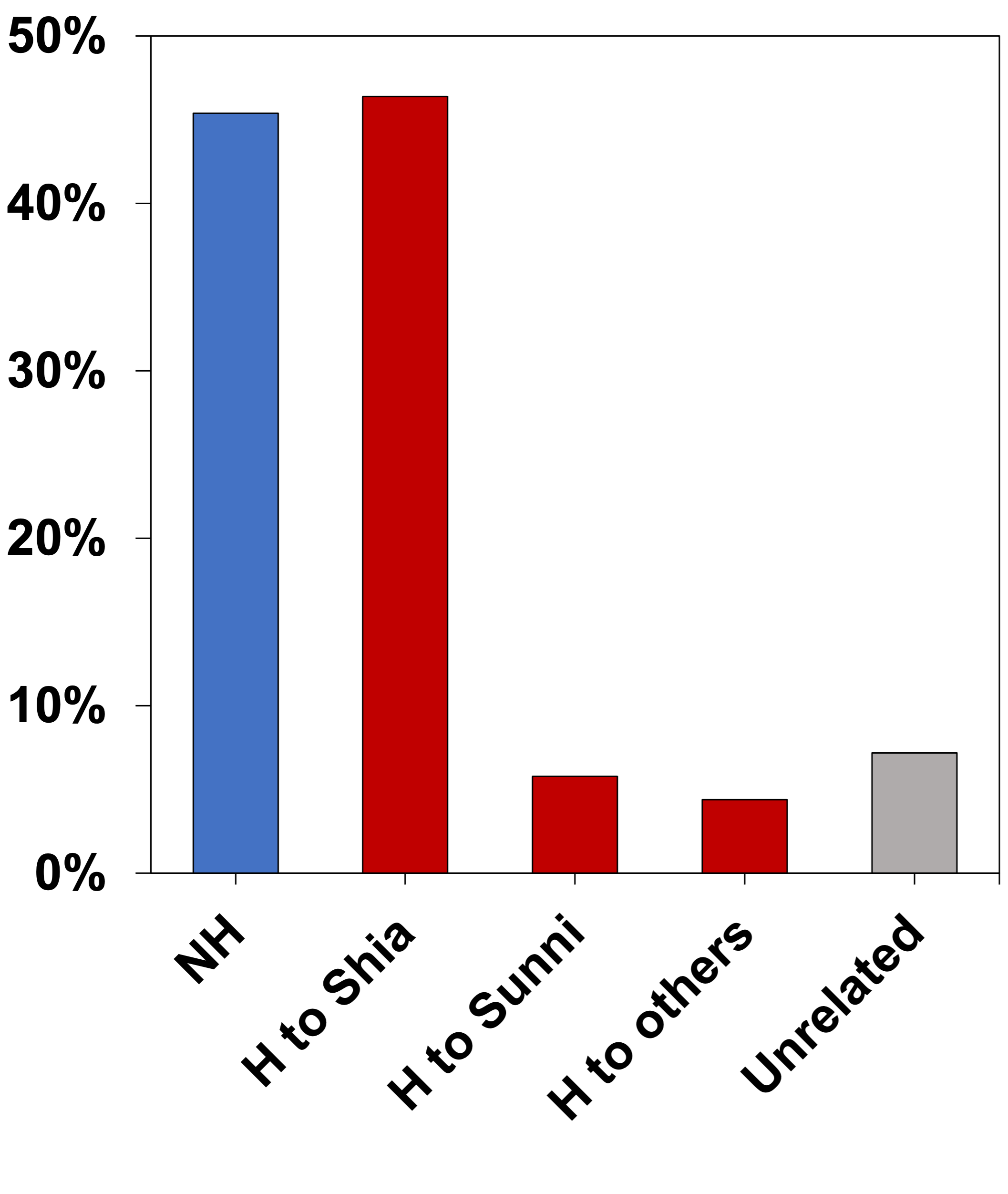}
\label{Shia}}
\subfloat[Atheists dataset]{\includegraphics[width=0.33\linewidth]{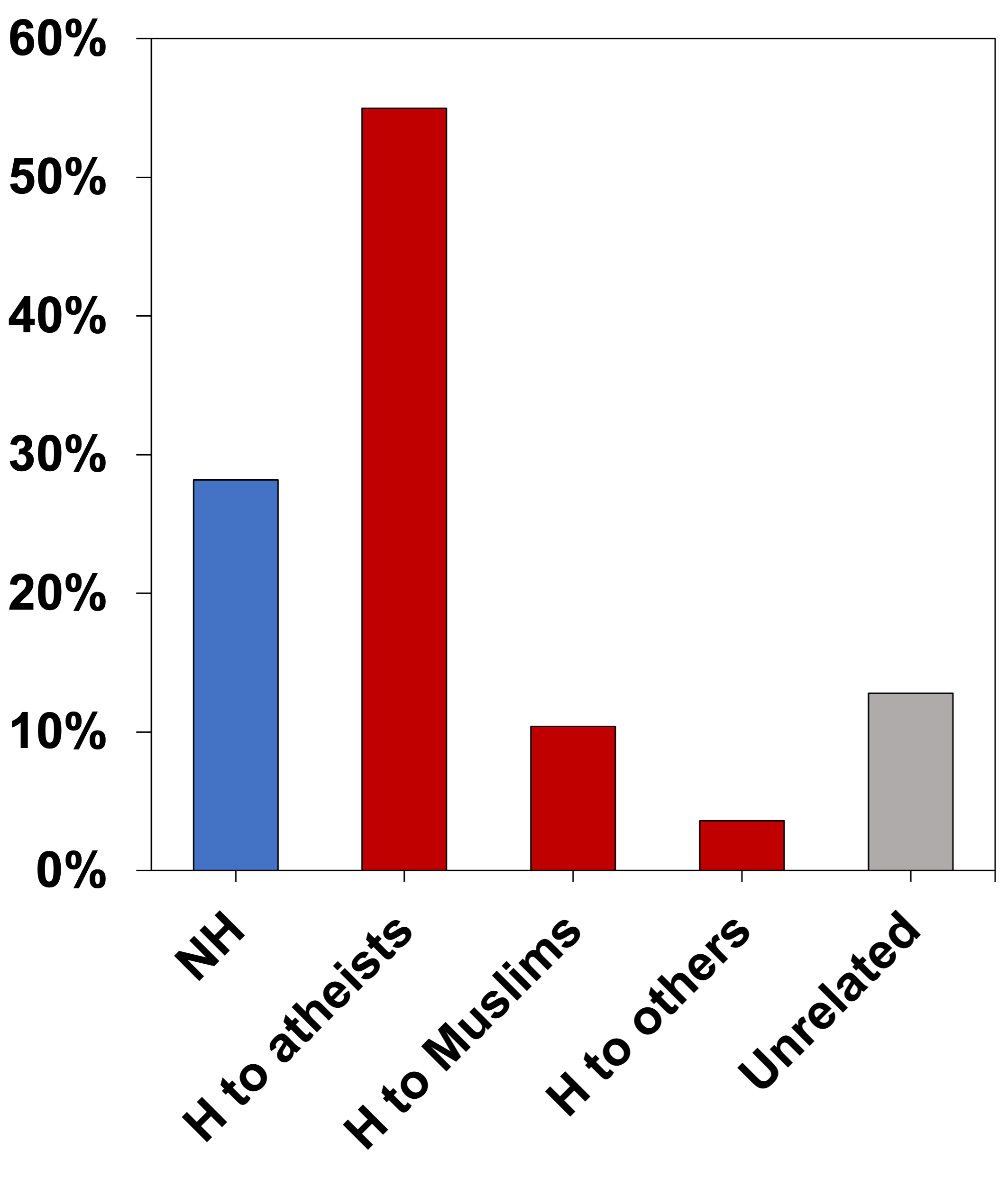}
\label{Atheists}}
\subfloat[Christians dataset]{\includegraphics[width=0.33\linewidth]{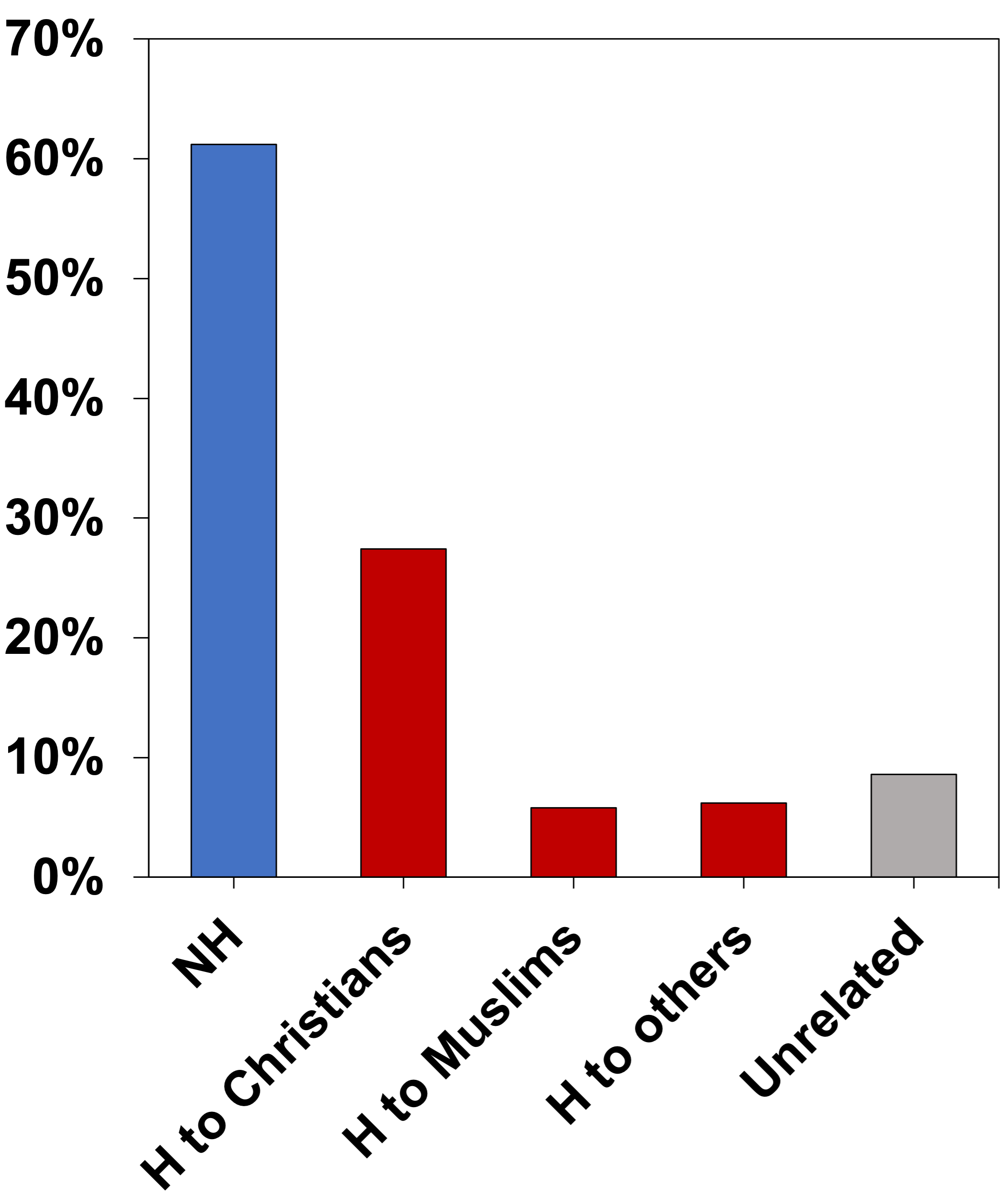}\label{Christians}}%
\medskip

\subfloat[Jews dataset]{\includegraphics[width=0.33\linewidth]{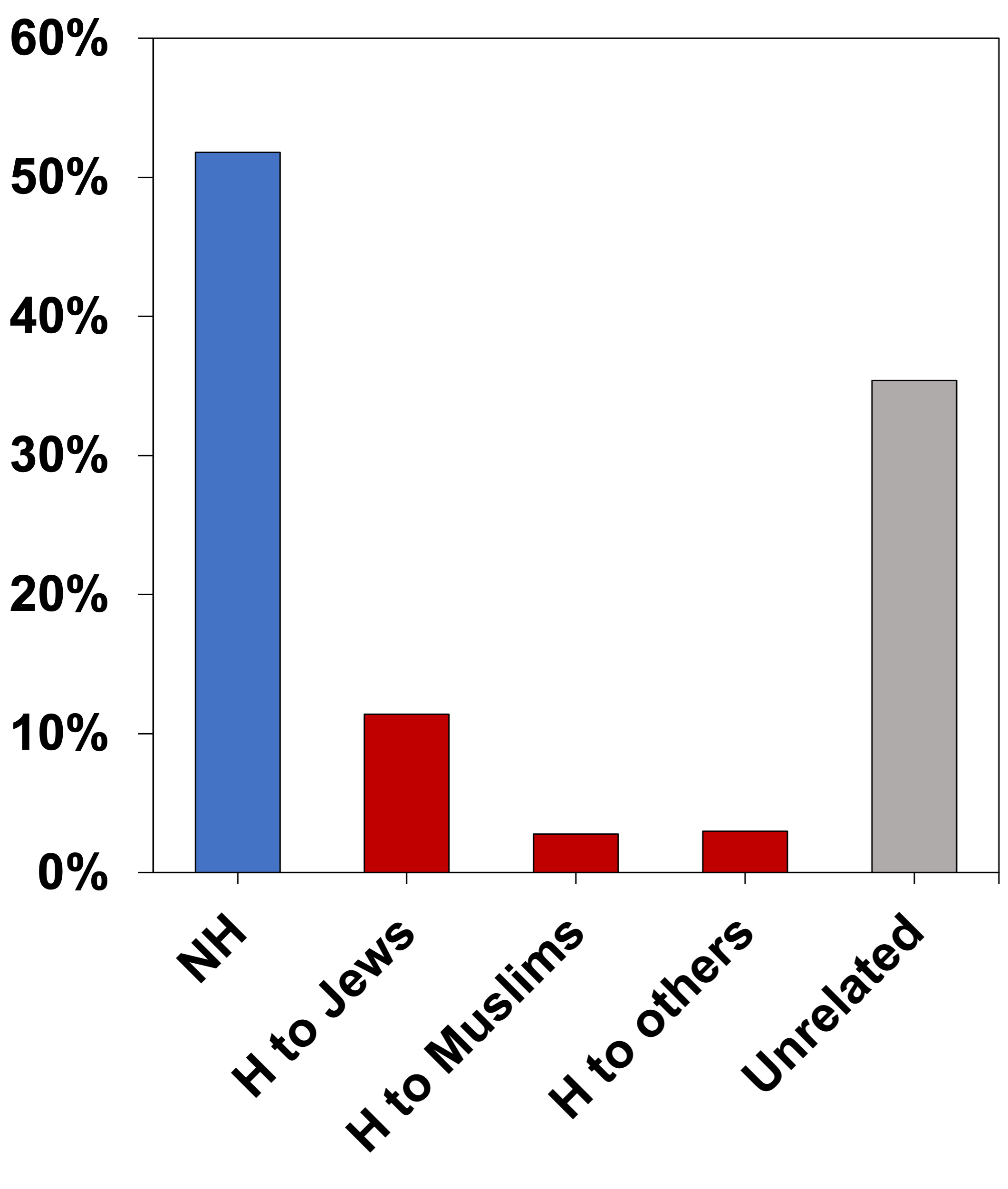}\label{Jews}}%
\subfloat[Muslims dataset]{\includegraphics[width=0.33\linewidth]{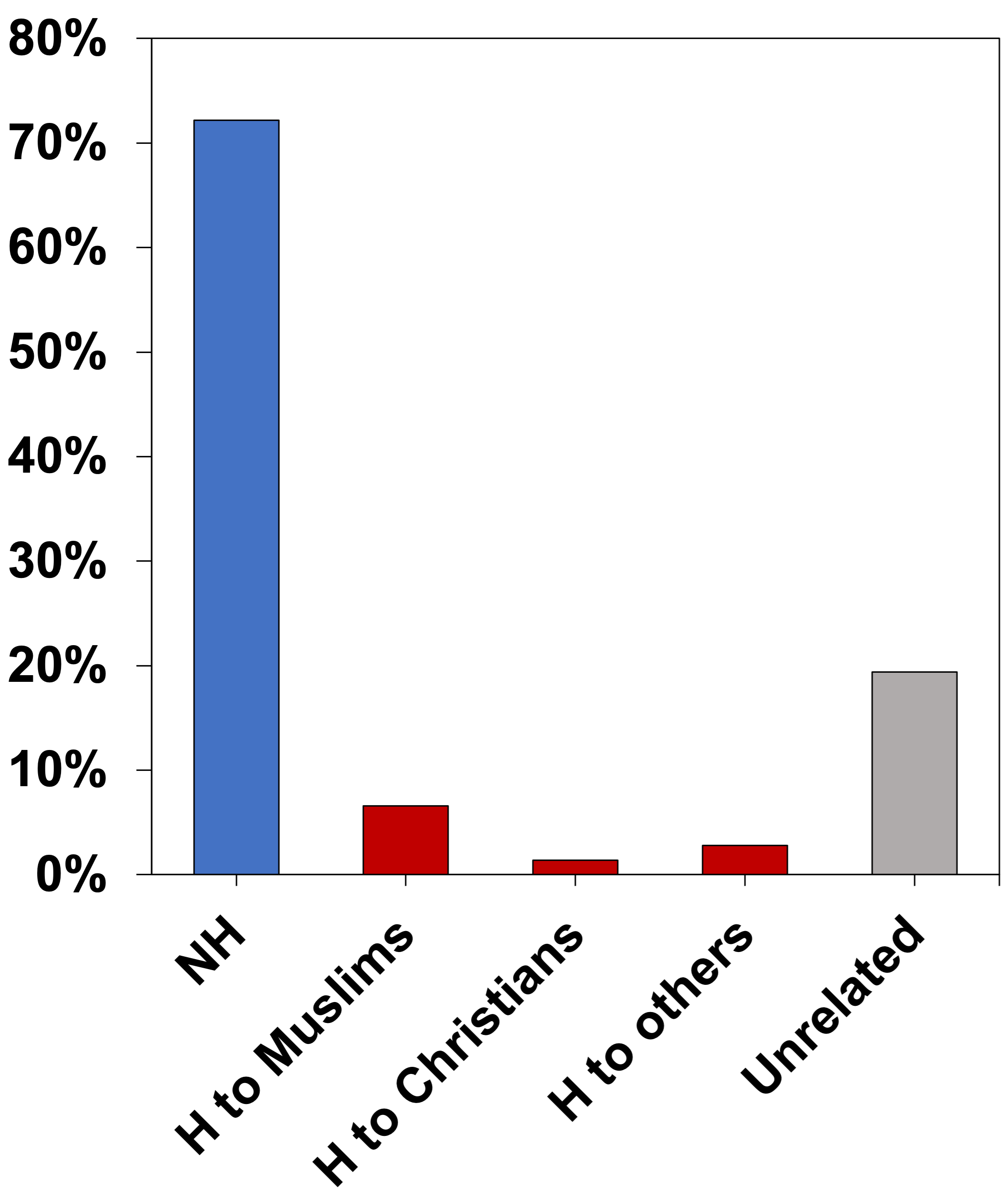}\label{Muslims}}
\subfloat[Sunni dataset]{\includegraphics[width=0.33\linewidth]{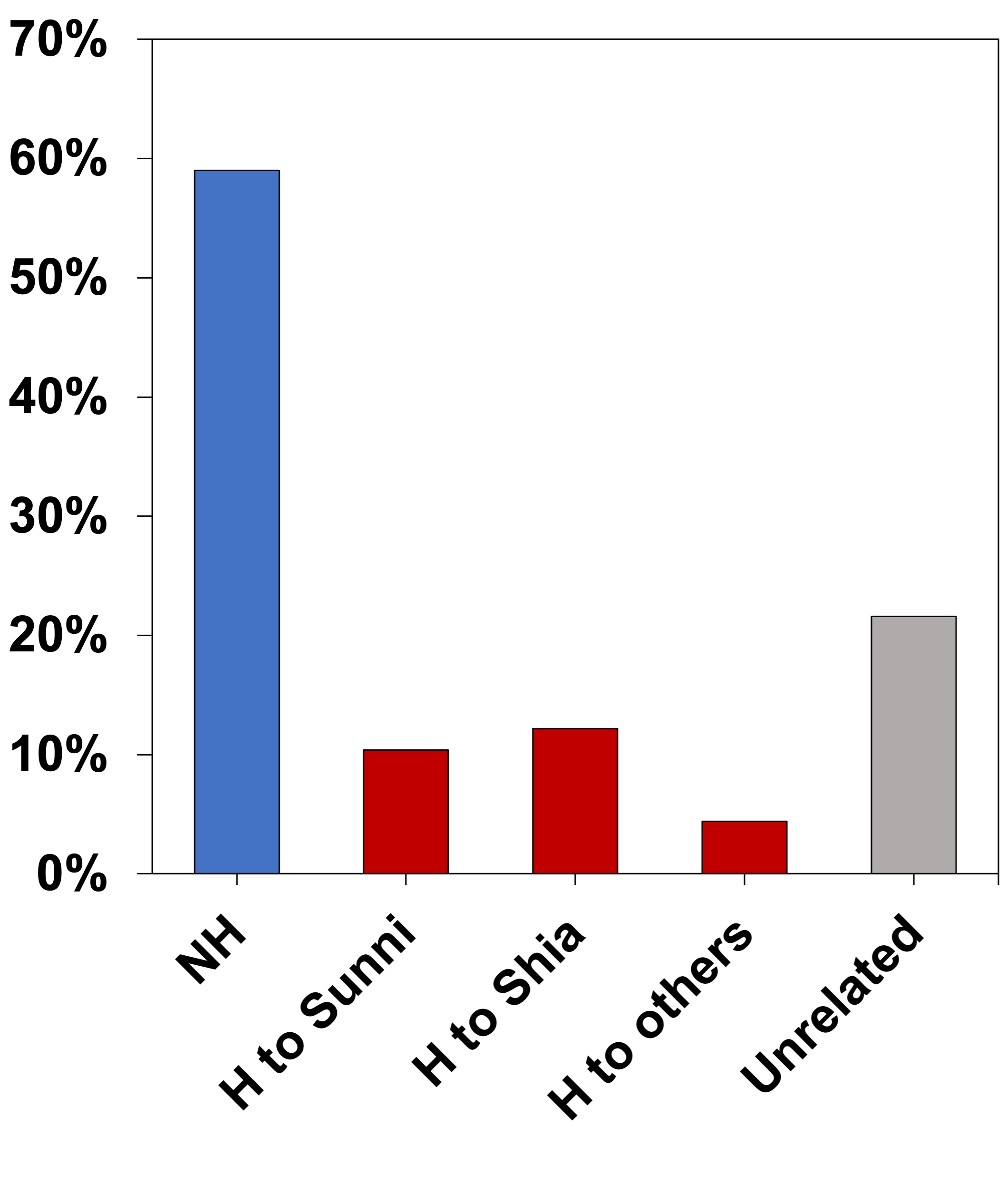}\label{Sunni}}
\setlength{\belowcaptionskip}{-12pt}
\caption{Distribution of annotations across religious-group datasets. (H) denotes hateful and (NH) denotes non-hateful.}
\label{annotations_across_religions}
\end{figure}

To get further insight, we looked at each of the 500 videos collected for each religious group individually (see Figure \ref{annotations_across_religions}). We observed that for the atheist dataset, more than half of the videos (55\%) were deemed hateful toward them; in the Shia dataset, almost half of the videos (46\%) were considered hateful toward them. Across all datasets, the most targeted religious group turned out to be the religious group for which the dataset was collected, except in the Sunni dataset, where the most targeted religious group was the Shia rather than the Sunni. 

\begin{figure}[!t]
  \centering
  \includegraphics[width=0.8\linewidth]{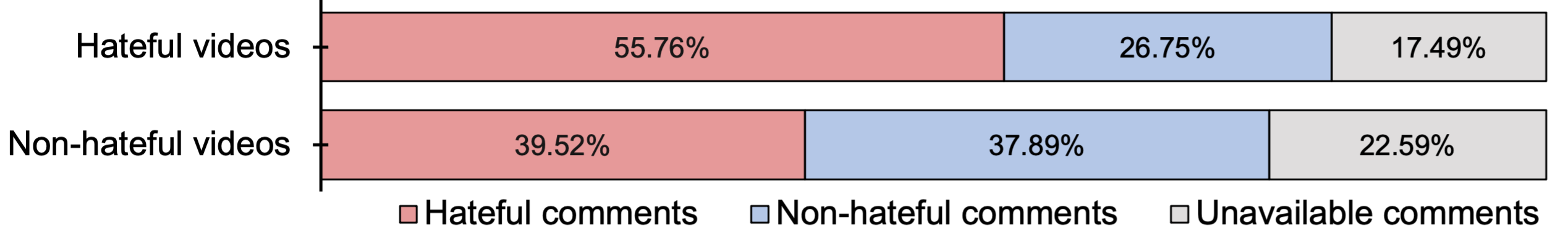}
  \caption{Distribution of video comment hate class for hateful and non-hateful videos. }
  \label{comments_distib}
\end{figure}

\begin{figure}[t]
\subfloat[Shia dataset]{\includegraphics[width=0.5\linewidth]{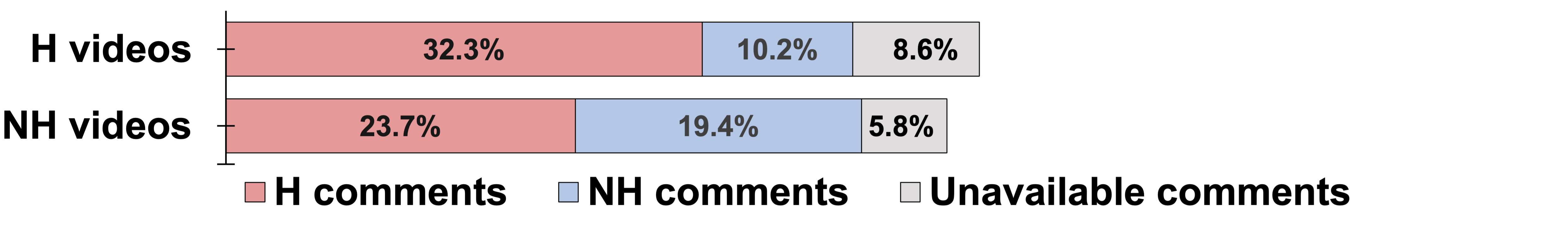}
\label{Shia}}
\subfloat[Atheists dataset]{\includegraphics[width=0.5\linewidth]{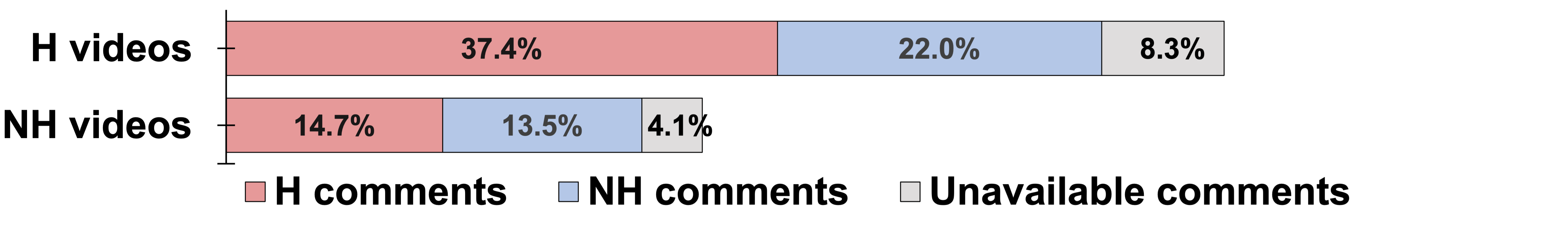}
\label{Atheists}}

\subfloat[Christians dataset]{\includegraphics[width=0.5\linewidth]{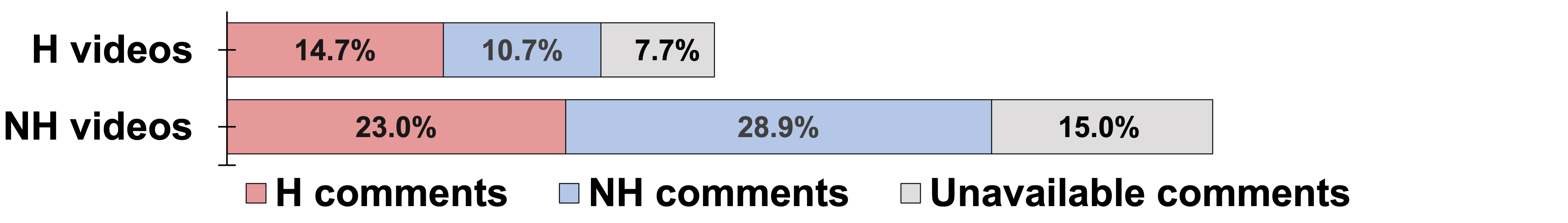}\label{Christians}}%
\subfloat[Jews dataset]{\includegraphics[width=0.5\linewidth]{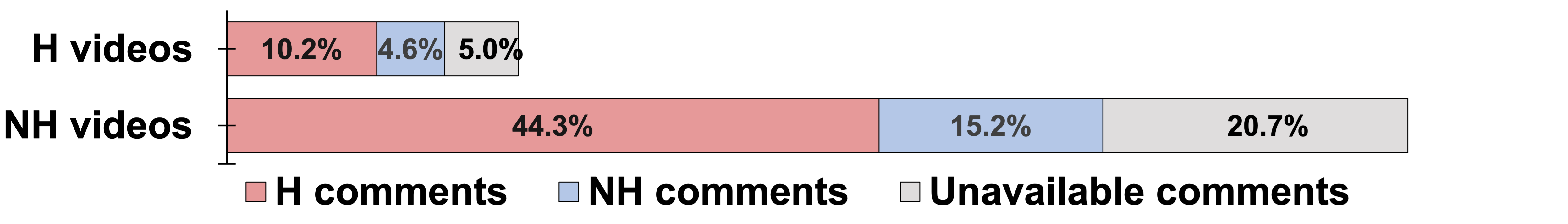}\label{Jews}}%

\subfloat[Muslims dataset]{\includegraphics[width=0.5\linewidth]{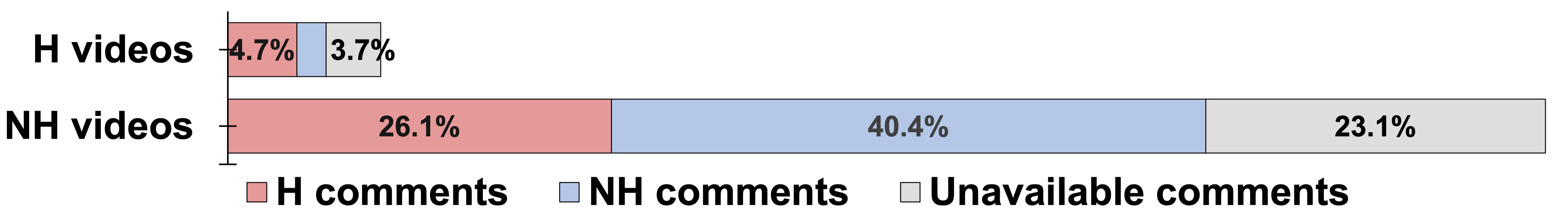}\label{Muslims}}
\subfloat[Sunni dataset]{\includegraphics[width=0.5\linewidth]{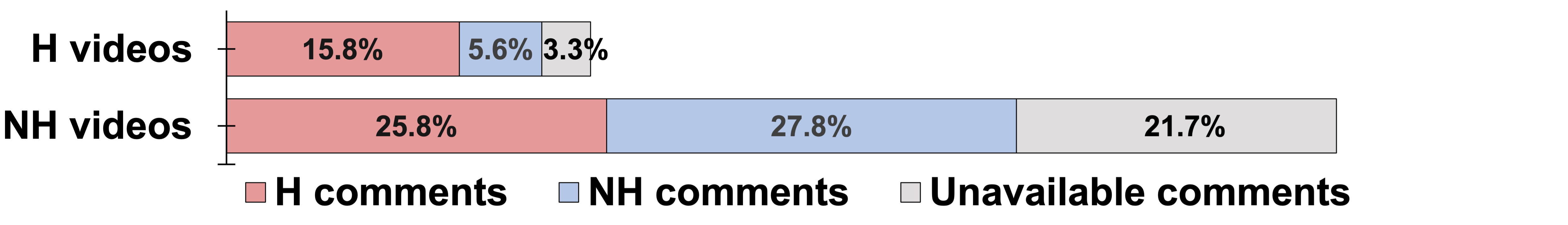}\label{Sunni}}
\caption{Distribution of video comment hate class for hateful and non-hateful videos across religious-group datasets. (H) denotes hateful, (NH) denotes non-hateful. }
\label{comments_by_relegion}
\end{figure}

Next, we investigate whether hateful videos had a significantly larger proportion of hateful comments than non-hateful videos. To get annotations for YouTube comments, we leveraged the Twitter hate speech classifier \cite{albadi2018they} that was trained on Arabic tweets discussing the same religious groups considered in this paper. We preprocessed YouTube comments following the same preprocessing methodology that was used in training the Twitter classifier. The Twitter classifier can handle text of length up to 50 words. Thus, we truncated YouTube comments to that length. This wasn't an issue since YouTube comments had an average length of 14 words and a maximum of 1,707 words. For each video, we applied the Twitter classifier on up to 100 of its recent comments and found the most common class (hateful vs. non-hateful) among its comments. Figure \ref{comments_distib} illustrates the distributions of hateful, non-hateful, and unavailable (i.e., disabled) comments for both hateful and non-hateful videos. As expected, hateful videos had larger proportions of hateful comments than non-hateful videos. This difference in distributions was found to be statistically significant, ${\chi}^2$ (2, $N$ = 2475) = 61, $p$ < 0.001. To gain deeper insights, we looked at comment hate class distributions for hateful and non-hateful videos across each religious group individually (see Figure \ref{comments_by_relegion}). Across all religious groups, hateful videos had larger proportions of hateful comments than non-hateful comments. As for non-hateful videos, we can see that for Jews, atheists, and Shia the number of non-hateful videos with hateful comments exceeded the number of non-hateful videos with non-hateful comments, which is counterintuitive. This difference was the widest for the Jews dataset, i.e., non-hateful Jews-related videos had the highest proportions of hateful comments, even higher than other religious group hateful videos. 


\section{Detection of Videos Promoting Religious Hate}
Building on our initial analysis reported in the previous section, we describe in this section our approach to identifying Arabic YouTube videos promoting religious hate using deep learning methods. We detail the features used, describe the model architecture, and report performance metrics on a hold-out testing dataset. 

\subsection{Methods}
We present a binary classification task to identify Arabic YouTube videos promoting religious hate. We focus on deep learning methods given their superiority and efficiency in different, but related classification tasks on YouTube \cite{papadamou2020disturbed,Kostantinos2022justAflue}. To train, validate and evaluate our model, we used the videos identified as either hateful or non-hateful (2,475 videos out of the ground truth dataset of 3,000 videos by excluding unrelated videos). We used 70\% of the data for training, 10\% for validation, and 20\% for testing. 

We implemented our model using Keras \cite{chollet2015keras}, a Python library for developing deep learning models, and TensorFlow \cite{abadi2016tensorflow} backend. This model was trained on different combinations of features that include title, description, tags, thumbnail, and statistical features (See the next subsection). We used Keras functional API as it allows for the handling of multiple inputs. We used Adam algorithm for optimization, cross entropy for the loss function, and trained our model in batches of size 32 to achieve optimal performance and stability. Finally, we evaluated our models based on precision, recall, F\textsubscript{1} score, accuracy, and area under the receiver operating characteristic (AUROC).

\subsection{Features Description and Model Architecture}

Below we describe the features we considered when developing our deep learning classifiers and explain their preprocessing steps.

\textbf{Title.} To transform video title words to word embeddings, we used an Arabic pre-trained word embedding model, AraVec 3.0 \cite{soliman2017aravec}. AraVec's \textit{Twitter-CBOW} model was trained on 67M Arabic tweets, and it has a vocabulary size of 1.5M words and an embedding dimensionality of 300. We preprocessed the title following the same preprocessing steps used by Aravec to maximize the number of found word embeddings for our title vocabulary. The size of the title vocabulary was 7,900 words, 87\% of these had an AraVec word embedding vector. The maximum title length after preprocessing was 23 words, and the average was 10 words. We padded the title feature vector to the maximum title length. 

\textbf{Description.} We preprocessed the video description similar to how we preprocessed the video title. However, unlike video title, video description tends to be lengthy, and thus we removed stop words from them. The maximum description length after preprocessing and stop words removal was 922 words, and the average was 47 words. The size of the description vocabulary was 31,444 words, 69.5\% of these were matched against AraVec vocabulary. To not confuse our model with extremely large description word count, we truncated/padded the description feature vector to the average description length. 

\textbf{Thumbnail.} We considered two methods for extracting thumbnail features. The first one was using transfer learning \cite{bozinovski2020reminder} through a pre-trained convolutional neural network (CNN) model. For that purpose, we used \cite{papadamou2020disturbed}'s thumbnail feature extractor module that internally uses Inception-v3 \cite{DBLP:journals/corr/SzegedyVISW15}, a pre-trained CNN model trained on millions of images from the ImageNet dataset to transform each thumbnail image to a meaningful  2048-dimensional feature vector. The model expects images to have a dimensionality of 299 $*$ 299 $*$ 3, and thus we down-scaled our thumbnail images from 360 $*$ 480 $*$ 3 to the required dimensionality. We also considered building our own CNN model to extract thumbnail features. However, the model using the pre-trained CNN model substantially outperformed the one using our own CNN; thus, we decided to use a pre-trained CNN model. 

\textbf{Tags.} Video tags are descriptive keywords optionally provided by the video uploader to provide context about the video content and are used to rank videos in search results. Similar to the title and description, we used AraVec word embedding model to transform tags into word embeddings. The maximum number of tags was 93, and the average was 21. The size of the tags vocabulary was 10,398 words, 74.33\% of these had a match against AraVec vocabulary. We padded the tags feature vector to the maximum tags length. 

\textbf{Statistical Features.} We considered the following statistical features: video view count, video like count, video dislike count, video duration in seconds, video comment count, channel view count, channel subscriber count, and channel video count. We also considered whether a given video had a larger proportion of hateful or non-hateful comments, which we were able to discern by applying the Twitter religious hate speech classifier, as discussed in Section \ref{groundtruth}. This resulted in having a statistical feature vector of length 11. We normalized the statistical feature vector so that all features ranged from  0 to 1 to prevent features that tend to have higher values (e.g., view count) from having more influence on the model weights learning process. 

\subsection{Model Architecture}

\begin{figure}[!t]
  \centering
  \includegraphics[width=\linewidth]{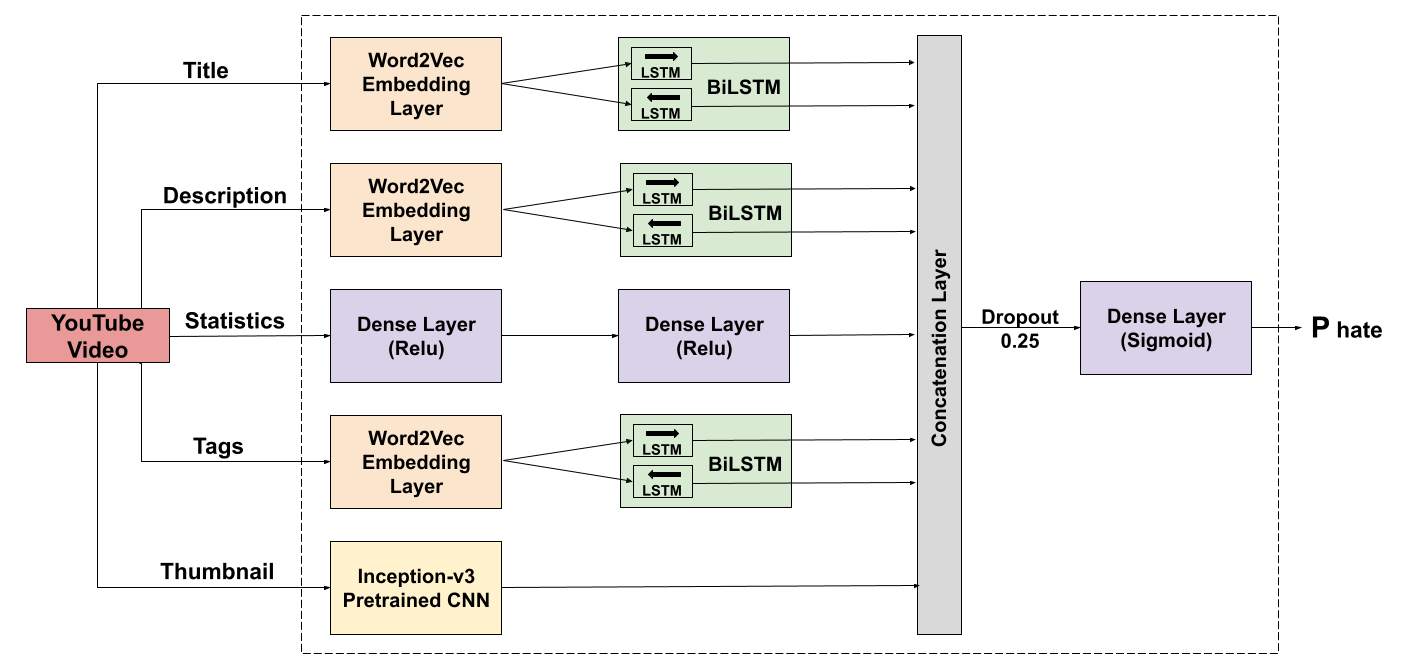}
  \caption{Model architecture of our deep learning classifier. }
  \label{model}
\end{figure}

Figure \ref{model} illustrates the architecture of our deep learning classifier. The classifier handles each feature type (i.e., title, description, tags, thumbnail, and statistics) in a separate branch. Title, description, and tags features are processed in architecturally similar branches consisting of: 1) a trainable embedding layer that maps words to their corresponding pre-trained word vectors; 2) a bidirectional long/short-term memory (BiLSTM)\cite{schuster1997bidirectional} layer with 240 units to capture wide-range contextual information from both preceding and succeeding words. The thumbnail is processed by a pre-trained CNN (Inception-v3) that transforms each thumbnail image into a 2048-dimensional feature vector. The statistical feature vector is fed into a fully connected dense layer with 64 units, followed by a \textit{ReLU} activation. The output of that is feed into another fully connected dense layer with 32 units, followed by a \textit{ReLU} activation. The outputs from all branches are then concatenated and regularized with a 0.25 dropout layer before making a final prediction using a \textit{sigmoid} activated, one-unit dense layer. 

\subsection{Results}
\label{classifier_results}
We developed several deep learning classifiers to identify hateful Arabic videos using different sets of features. Table \ref{EvaluationResults} reports performance metrics of these classifiers on a hold-out testing dataset. At first, we investigated model performances trained on each feature type individually. Among individual features, the highest F\textsubscript{1} score (0.64) was achieved when training the model on the title feature, followed by the description feature (0.53). Next, we explored different combinations of textual features and found that training the model on both title and description features increased F\textsubscript{1} score by 3 points and enhanced recall by 10 points. Finally, to further improve the model performance, we explored training the model on the thumbnail and statistical features in addition to the best textual features (i.e., title and description). As illustrated in Table  \ref{EvaluationResults}, training the model on title, description, and statistical features provided the best F\textsubscript{1} score (0.69) across all combinations of features. We use this classifier for all of our upcoming analyses. 

To our knowledge, our classifier represents the first effort in detecting hateful Arabic content on YouTube, and thus we couldn't directly compare our model performance to any other model. However, there are other models that were developed to detect different but related problematic content on YouTube. For example,  the researchers in \cite{Kostantinos2022justAflue} developed a deep learning classifier that was able to detect pseudo-scientific YouTube videos with a 0.74 F\textsubscript{1} score. In \cite{mariconti2019you}, the authors developed a detection model to identify YouTube videos that are being targeted by third-party coordinated hate attacks with an F\textsubscript{1} score of 0.46. Impressively, the binary classifier reported in \cite{kaushal2016kidstube} was able to detect inappropriate videos targeting kids with an F\textsubscript{1} score of 0.82. Our model delivers comparable performance to these models and can effectively detect YouTube videos promoting religious hatred with an accuracy of 0.76. However, we acknowledge that this is not a perfect performance, and it reflects the subjectivity and nuance in recognizing hate speech. It also signals the need for more research efforts in this area.

\begin{table}[t]
  \caption{Evaluation results of our classification models using different features. Models with the highest F\textsubscript{1} score within each feature group are highlighted in bold. }
  \label{EvaluationResults}
  \begin{tabularx}{\linewidth}{llccccc}
    \toprule
    \textbf{Feature group} & \textbf{Features} &  \textbf{P} &  \textbf{R} & \textbf{F\textsubscript{1} } &  \textbf{Acc.} & \textbf{AUC} \\
    \midrule
    \multirow{5}{*}{\pbox{1.9 cm}{Individual features}} 
        & \textbf{Title}       & \textbf{0.68} & \textbf{0.61} & \textbf{0.64} & \textbf{0.76} & \textbf{0.8} \\
        & Description & 0.51 & 0.56 & 0.53 & 0.62 & 0.67 \\
        & Tags        & 0.56 & 0.31 & 0.4  & 0.66 & 0.65 \\
        & Thumbnail   & 0.42 & 0.52 & 0.46 & 0.58 & 0.59 \\
        & Statistics  &	0.57 & 0.28 & 0.37 & 0.67 & 0.63 \\
    \midrule
    \multirow{3}{*}{\pbox{2cm}{Textual features}}
        & \textbf{Title+description} & \textbf{0.64} & \textbf{0.71} & \textbf{0.67} & \textbf{0.76} & \textbf{0.81}  \\
        &Title+tags        & 0.65 & 0.65 & 0.65 & 0.75 & 0.8  \\
        &Title+description+tags & 0.61 & 0.69 & 0.65 & 0.73 & 0.78 \\
    \midrule
    \multirow{3}{*}{\pbox{2cm}{Best textual features + other features}}
    & Title+description+thumbnail & 0.64 & 0.67 & 0.65 & 0.73 & 0.8 \\
    & \textbf{Title+description+statistics} & \textbf{0.67} & \textbf{0.71} & \textbf{0.69} & \textbf{0.76} & \textbf{0.79} \\
    & Title+description+thumbnail+statistics & 0.64 & 0.56 & 0.6 & 0.75 & 0.81 \\
  \bottomrule
\end{tabularx}
\end{table}

Given that our classifier was trained and tested on search result videos, we next validate our classifier on recommended videos. To do that, we randomly selected 100 recommended videos (20 videos from each recommendation level). To obtain ground truth for these videos, the first two authors reviewed these videos and came to consensus on their label, following the hate speech definition and criteria discussed in Section \ref{sec:annotation}. We make public the list of videos used in this validation step along with the true label that we assigned to the video and the predicted label assigned by the classifier \footnote{\url{https://osf.io/cf9w8/?view_only=aa81f43ff28c4faaa7514ccccc6a386c}}. Table \ref{performanc_recommended} summarizes the performance of our classifier on level 1 to level 5 recommended videos. Due to the small sample size for each recommendation level, the model performance is highly sensitive to the selected videos in each level. For example, in level 5 recommendation, precision and recall are zero, and that is due to not having any true examples of hateful videos in the sample for that level.

Overall, the classifier on recommended videos delivered comparable results to those on search result videos with only 2 points down in F\textsubscript{1} score. While the precision decreased by 7 points, the recall improved by 4 points. This indicates that our model detects most hateful videos with the cost of flagging some innocuous videos as hateful. We argue that in the case of detecting hateful videos, recall weighs heavier than precision as the goal is to detect all videos that could potentially be hateful and then have human moderators review these videos for a final decision. Additionally, the probability that our classifier would incorrectly classify a non-hateful video as hateful (i.e., false positive rate) is only 0.04. It is also worth noting that 50\% of misclassified videos across the levels had a predicted hate probability of around 0.5, i.e., they were considered edge cases for the classifier. Given these relatively good results, we believe that our classifier can be reliably used to answer our research questions.

\begin{table}[t]
  \caption{Classifier performance on recommended videos.}
  \label{performanc_recommended}
  \begin{tabular}{lcccccccc}
    \toprule
    \textbf{Recommendation level}  & \textbf{P} &  \textbf{R} & \textbf{F\textsubscript{1} } &  \textbf{Acc.} & \textbf{TP} &  \textbf{FN} &  \textbf{FP}  &  \textbf{TN} \\
    \midrule
    Level 1 & 1 & 0.67 & 0.8 & 0.95 & 2 & 1 & 0 & 17 \\
    Level 2 & 0.5 & 0.5 & 0.5 & 0.9 & 1 & 1 & 1 & 17 \\
    Level 3 & 0.67 & 1 & 0.8 & 0.95 & 2 & 0 & 1 & 17 \\
    Level 4 & 0.5 & 1 & 0.67 & 0.95 & 1 & 0 & 1 & 18 \\
    Level 5 & 0 & 0 & 0 & 0.95 & 0 & 0 & 1 & 19 \\
     \midrule
    All levels & 0.6 & 0.75 & 0.67 & 0.94 & 6 & 2 & 4 & 88 \\
  \bottomrule
\end{tabular}
\end{table}

\section{Analysis}
In this section, we answer the first two research questions concerning getting proxy indicator of the spread of hateful videos on YouTube and the platform's tendency to recommend hateful videos. Confirmed by a normality test, our data is not normally distributed, and thus we use Chi-squared test \cite{Pearson1900} for our comparisons. 


\subsection{The Extent of Religiously Intolerant Videos (RQ1)}
\label{prevalence}
To assess the prevalence of religiously intolerant Arabic videos, we applied our deep learning classifier to over 350K unique recommended videos collected through five levels of recommendations. Table \ref{level_dist} illustrates the proportions of hateful and non-hateful videos in each level of recommendations. These proportions represent unique videos (no duplicates) within each level of recommendations. For ``all levels'' entry, we also dropped duplicated videos collected through multiple levels of recommendations. We found that level 1 recommendations contained the highest proportions of hateful videos (about 21\%). Hateful videos tended to decrease as we moved deeper into the recommendation graph, reaching about 12\% in level 5 recommendations. We viewed random samples from each recommendation level to find a possible explanation for this diminishing trend in hateful videos. We observed that the further we moved in the recommendation graph, the more videos there were that were unrelated to religious discussions. Overall, 12\% of the 351,262 unique recommended videos across all levels were hateful. \textbf{Insight:} \textit{YouTube recommendations collected by our study contain a significant proportion of religiously intolerant content, particularly prevalent in level 1 recommendations.}

\begin{table}[t]
  \caption{Hateful and non-hateful video distributions within each level of YouTube's recommendation graph.}
  \label{level_dist}
  \begin{tabular}{ccc}
    \toprule
    Level \# & Non-hateful (\%) & Hateful (\%)\\
    \midrule
    Level 1 & 6,391 (79.20\%) & 1,678 (20.80 \%) \\ 
    Level 2 & 19,199 (82.33\%) & 4,120 (17.67 \%) \\ 
    Level 3 & 51,192 (84.41\%) & 9,454 (15.59 \%) \\ 
    Level 4 & 119,001 (86.36\%) & 18,788 (13.64 \%) \\ 
    Level 5 & 252,086 (88.08\%) & 34,104 (11.92 \%) \\ 
     \midrule
    all levels & 308,552 (87.85\%) & 42,710 (12.15 \%)\\ 
  \bottomrule
\end{tabular}
\end{table}

\subsection{YouTube's Tendency to Recommend Hateful Videos (RQ2)}

We followed the methodology described in \cite{papadamou2020disturbed} to measure the tendency of YouTube to recommend hateful videos. We leveraged 929,596 recommendations collected through traversing five levels of recommendations to create a directed graph that resembled YouTube recommendations in which nodes represent videos and edges represent a recommendation activity. We computed the out-degree in terms of hateful (H) and non-hateful (NH) videos for each node. Then, we counted the number of transitions the graph made between all different combinations of hate classes: NH to NH, NH to H, H to NH, and H to H. As illustrated in Table \ref{graph_transitions}, we found that recommendations pointing to hateful videos (NH to H and H to H) comprised about 15\% of overall recommendations . In particular, we note that 12\% of recommendations to non-hateful videos are hateful. Additionally, about 31\% of videos recommended to hateful videos are also hateful. Note that the number of unique recommended videos is 351,262, while the total number of transitions is 929,596 which includes transitions to duplicate videos (i.e., same videos that YouTube recommended in multiple instances). \textbf{Insight:} \textit{Our recommendation graph analysis offers preliminary evidence on the possible radicalization role played by YouTube recommendations which could expose and/or reinforce exposure of users to hateful videos and possibly isolate them in filter bubbles of extreme content.}


\begin{table}[t]
  \caption{The distribution of transitions between hateful (H) and non-hateful (NH) videos for all recommendation levels.}
  \label{graph_transitions}
  \begin{tabular}{lc}
    \toprule
    Transition type & Number of transitions (\%)\\
    \midrule
    NH $\rightarrow$ NH & 695,381  (74.8 \%) \\ 
    NH $\rightarrow$ H  & 94,567  (10.17 \%) \\ 
    H  \hspace{0.16cm}  $\rightarrow$  NH & 96,858  (10.42 \%) \\ 
    H \hspace{0.16cm} $\rightarrow$  H & 42,790      (4.61 \%) \\ 
  \bottomrule
\end{tabular}
\end{table}

\section{Personalized Auditing of YouTube's Algorithm (RQ3)}
\label{personalization_audits}
In this section, we carry out an audit assessment to measure the effect of personalization based on religious ideology, Islamic denomination, and gender on the extent of exposure to hateful content.
\subsection{Methods}
\label{Methods}

\textbf{Creating User Profile.} To understand the impact of personalization on YouTube algorithms, our first task was to create several user profiles that differ from one another in some specific personal attributes. For this study, we considered binary values of the following personal attributes: 1) religious ideology (radical vs. moderate); 2) Islamic denomination (Sunni vs. Shia); 3) gender (female vs. male). We carefully crafted eight Google accounts, each with a distinctive set of personal attributes: (1) Radical, Sunni, Male; (2) Radical, Sunni, Female; (3) Radical, Shia, Male; (4) Radical, Shia, Female; (5) Moderate, Sunni, Male; (6) Moderate, Sunni, Female; (7) Moderate, Shia, Male; (8) Moderate, Shia, Female. We set the age for all eight accounts to 26 as this was the reported average age of individuals who have been charged with ISIS-related activities in the United States \cite{ISISInAmerica}. 


To establish the Islamic denomination and religious ideology attributes, we first needed to identify Islamic clerics from both denominations known to promote either radical or moderate ideology. To do this, we interviewed five members of the Shia (3 males and 2 females) and six members of the Sunni (2 males and 4 females). Given the highly sensitive topic of the interview, those members were recruited from the authors personal network. We asked them to provide names of Islamic clerics from their denomination whom they believe promote coexistence, and acceptance values (i.e., moderate clerics) and those promoting jihad and other extreme ideologies (i.e., radical clerics). We only considered clerics who got mentioned twice or more for the same type of religious ideology. Note that these interviews were conducted to only get an initial list of clerics. Afterward, we verified that these clerics indeed support radical or moderate ideology by watching/reading some of their videos/books while paying attention to how they regard members of other religious groups, i.e., whether they were demonizing or tolerating other religions. Radical Sunni clerics in our list, in particular, publicly supported militant jihad and the death penalty for apostates. Overall, we identified ten radical and nine moderate Sunni clerics and two radical and five moderate Shia clerics. Interestingly, all five Shia interviewees agreed on the same two radical clerics. Due to the sensitivity of the topic, we will not be sharing the names of mentioned clerics.

\textbf{Building Watch History.} After identifying Sunni and Shia clerics who were at either end of the Islamic spectrum, we started building the watch history for each of the eight profiles. For example, the watch history for a \textit{radical Sunni} profile was created by watching videos of various \textit{Radical Sunni} clerics. The watching took place four days a week, for a total of nine weeks. To control for temporal effects, we randomly selected the four days of the week, which included both weekdays and weekends. The time of the day the watching took place was also chosen randomly and had mornings, afternoons, and evenings. The arrangement of the profiles (based on religious ideology and Islamic denomination) over the days of the week was also randomized. Finally, we randomly distributed corresponding cleric names over the weeks. 

The watching for the male (by the 1st author) and female (by the 2nd author) profiles of the same Islamic denomination and religious ideology happened simultaneously, and each watching session lasted for 30 minutes. To reduce the possibility of confounding factors biasing our audit assessment, all watching was conducted at the same geographic location through devices with the same specifications and connected to the same WiFi network. To eliminate noise from tracked cookies and browsing history, we browsed YouTube in private mode using Firefox. At the start of every watching session, we logged-in to the assigned profile and then searched for the same assigned Islamic cleric. There were some variations in search results between the male and female profiles, even with a clean watch history, and thus watching the same video at the same time by the two profiles was not possible. Note that our goal is to audit gender as a profile setting rather than having the profile behave in a gender-specific way. Thus, to minimize personal differences when selecting which videos to watch, both authors selected videos that were most relevant to the search query and with a high number of views. In total, we created 4.5 hours of watch history for each profile.


\subsection{Search and Recommendation Audits}
We conducted two systematic audit experiments to investigate whether personalization contributes to greater levels of exposure to hateful content in both search results (\textit{search audit}) and recommendations (\textit{recommendation audit}). 

\textbf{Search Audit.} The primary purpose of the search audit is to investigate the effect of personalization on the proportion of hateful videos returned in search results. The search keywords of interest are the Arabic equivalent of the words: Shia, Sunni, Jews, Christians, and atheists. For each profile, we logged in to YouTube using that profile's credentials. We then performed search queries using the aforementioned keywords and considered the top 10 search results for each keyword. To minimize accidental bias created by the order of profiles or searched keywords, the following measures were taken: 1) the user profile and keyword selection were randomized; 2) we kept at least 11 minutes of interval between consecutive search to minimize the carry-over effect  \cite{hannak2013measuring}; 3) experiments were conducted in private mode using Firefox. We semi-automated the process of collecting video ids for each search result. Google was actively detecting logins using automated software, and thus fully automating the process (e.g., using Selenium bots) was not possible. Thus, we wrote a script that handled the randomization of profiles and keywords, processed HTML pages to identify the top 10 video ids, and enforced wait intervals between consecutive search queries. We manually handled the login process, typing of the keywords, and downloading of search result HTML pages. Overall, 400 video ids were collected during the search audit. 


\textbf{Recommendation Audit.} The recommendation audit aims to explore if there is a significant difference in the volume of hateful videos recommended to each profile. After completing the search audit, we logged in to each profile in private mode and collected the top 10 recommendations for all videos collected in its search audit. As with search audit, the logins, opening of video pages, and downloading of HTML pages were handled manually. The script randomly selected a profile, randomly selected a keyword, provided us with video ids collected for that keyword, and processed downloaded HTML pages to identify the top 10 recommendations for each video. For each profile, a total of 500 recommendations were collected, resulting in a total of 4,000 video ids collected during the recommendation audit.

After completing both audits, we used YouTube API to retrieve metadata for the collected videos.


\subsection{Results}
Here we conduct statistical analysis to assess the extent of biases exacerbated by the YouTube's algorithm based on personal attributes. We use Chi Square ${\chi}^2$ test to test statistical significance given that all observations are independent and all expected counts are larger than 10. As a measure of effect size, we use Relative Risk (RR) and 95\% confidence intervals. To adjust for multiple testing, we use Benjamini-Hochberg procedure \cite{benjamini1995controlling} to control false discovery rate (FDR) at level 0.05.


\begin{table}[t]
  \caption{Personal attributes with significant effects on recommended videos (FDR-survived results only) and their effect size reported using Relative Risk (RR) and 95\% confidence interval (95\% CI).}
  \label{effect_size}
  \begin{tabular}{lcclcc}
    \toprule
    Variable & NH &  H & Statistical test & RR & 95\% CI\\
    \midrule
    Religious ideology &  & & & & \\
    \hspace{0.3cm} Moderate Shia & 737 & 263  &  \multirow{2}{*}{\pbox{4cm}{${\chi}^2$(1,2000)=13.66, $p$<0.01}}  & \multirow{2}{*}{\pbox{4cm}{1.29}} & \multirow{2}{*}{\pbox{4cm}{1.13-1.48}}   \\
    \hspace{0.3cm} Radical Shia & 661 & 339 & & &  \\
    Islamic denomination & & & & & \\
    \hspace{0.3cm} Moderate Shia & 737 & 263 & \multirow{2}{*}{\pbox{4cm}{${\chi}^2$(1,2000)=7.44, $p$<0.05}} & \multirow{2}{*}{\pbox{4cm}{1.21}}  & \multirow{2}{*}{\pbox{4cm}{1.1-1.39}}  \\
    \hspace{0.3cm} Moderate Sunni & 682 &  318 & & \\
    Gender & & & & & \\
    \hspace{0.3cm} Sunni female & 715 & 285 & \multirow{2}{*}{\pbox{4cm}{${\chi}^2$(1,2000)=5.40, $p$<0.05}} & \multirow{2}{*}{\pbox{4cm}{1.16}} & \multirow{2}{*}{\pbox{4cm}{1.02-1.33}}  \\
    \hspace{0.3cm} Sunni male & 669 & 331 & &  & \\
    All recommendations & & & & & \\
    \hspace{0.3cm} Non-personalized  & 6,391 & 1,678 &  \multirow{2}{*}{\pbox{4cm}{${\chi}^2$(1,12069)=136, $p$ < 0.001}}& \multirow{2}{*}{\pbox{4cm}{1.46}} & \multirow{2}{*}{\pbox{4cm}{1.37-1.56}}  \\
    \hspace{0.3cm} Personalized  & 2,782 & 1,218 & & & \\
  \bottomrule
  \multicolumn{5}{l}{$p$-values are FDR-adjusted.} \\
\end{tabular}
\end{table}

\textbf{Religious Ideology Effect.} To investigate whether being at either end of the Islamic spectrum  affects the returned volume of hateful videos, we first compared the two high-level groups (the four profiles with radical attribute vs. the four profiles with moderate attribute). While the result was not significant for the search audit, it was close to being statistically significant for the recommendation audit at FDR-adjusted $p$=0.083. We then compared subgroups for religious ideology effect. We found that radical Shia profiles were recommended significantly more hateful videos (30\% more) than moderate Shia profiles, ${\chi}^2$(1,2000)=13.66, FDR-adjusted $p$<0.01 (refer to Table \ref{effect_size}). We also found that radical Shia profiles watching Sunni-related videos were recommended almost twice as much hateful videos compared to their moderate counterparts (${\chi}^2$(1,400)=7.68, FDR-adjusted $p$<0.05). As for Sunni profiles, we have not observed a religious ideology effect for them at any level. This suggests a clear distinction between radical and moderate Shia, while the boundaries seem to be not as clear between radical and moderate Sunni. \textbf{Insight:} {\it  There is a religious ideology effect on recommendations to Shia profiles, in which radical Shia profiles have 1.29 times the risk of getting recommended hateful videos compared to moderate Shia profiles.}


\textbf{Islamic Denomination Effect.} We explored whether being a member of an Islamic denomination (Shia vs. Sunni) affects the amount of hateful content presented to the users. Comparing all Shia vs. all Sunni profiles didn't yield any significant difference in their exposure to hateful content. However, when comparing subgroups for Islamic denomination effect, we found that moderate Sunni profiles were recommended significantly more hateful videos (21\% more) than moderate Shia profiles, ${\chi}^2$(1,2000)=7.44, FDR-adjusted $p$<0.05. The only other comparison that yielded marginally significant Islamic denomination effect (FDR-adjusted $p$=0.09) is between radical Shia profiles and radical Sunni profiles, in which radical Shia profiles were recommended more hateful videos. \textbf{Insight:} {\it There is an Islamic denomination effect on recommendations to moderate profiles, in which moderate Sunni profiles are 21\% more likely to be recommended hateful videos than moderate Shia profiles}.

\textbf{Gender Effect.} To study the effect of gender on returned search results and recommendations, we first compared all four male vs. all four female profiles. Results revealed no significant difference for both search and recommendation results. However, when comparing subgroups, the gender effect was only observed in videos recommended to Sunni profiles, ${\chi}^2$(1,2000)=5.40, FDR-adjusted $p$<0.05. \textbf{Insight:} {\it There is a gender effect on Sunni profiles, in which male Sunni profiles have 16\% increase in the risk of getting recommended hateful videos compared to female Sunni profiles}. 

\textbf{Personalization Effect.} To examine the effect of personalization, we compared results returned in the search and recommendation audits, which are affected by personalization, against the search and first-level recommendations returned using YouTube API (Section \ref{groundtruth} and \ref{prevalence}), which doesn't account for personalization. The percentage of hateful videos increased from 20.80\% (level 1 recommendation) to 30.45\% (recommendation audit). This difference in distribution between personalized and non-personalized recommendations was found to be statistically significant, ${\chi}^2$(1,12069)=136,  FDR-adjusted $p$ < 0.001. On the other hand, we didn't observe a significant difference between personalized and non-personalized search results. \textbf{Insight:} {\it Personalization in general increases the risk of getting recommended hateful videos by 46\%.}

\begin{figure}[t]
\subfloat[Recommendation audit]{\includegraphics[width=0.5\linewidth]{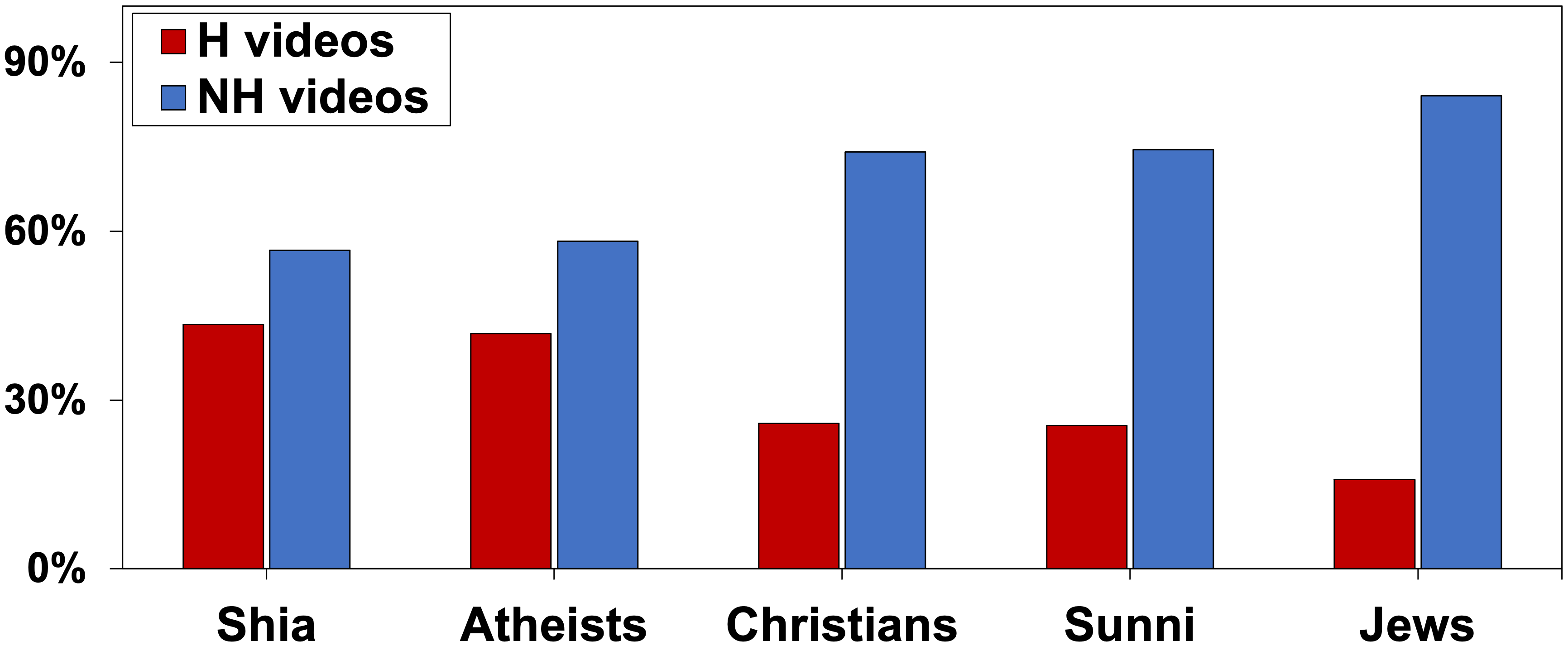}
\label{recom_keyword_dist}}
\subfloat[Search audit]{\includegraphics[width=0.5\linewidth]{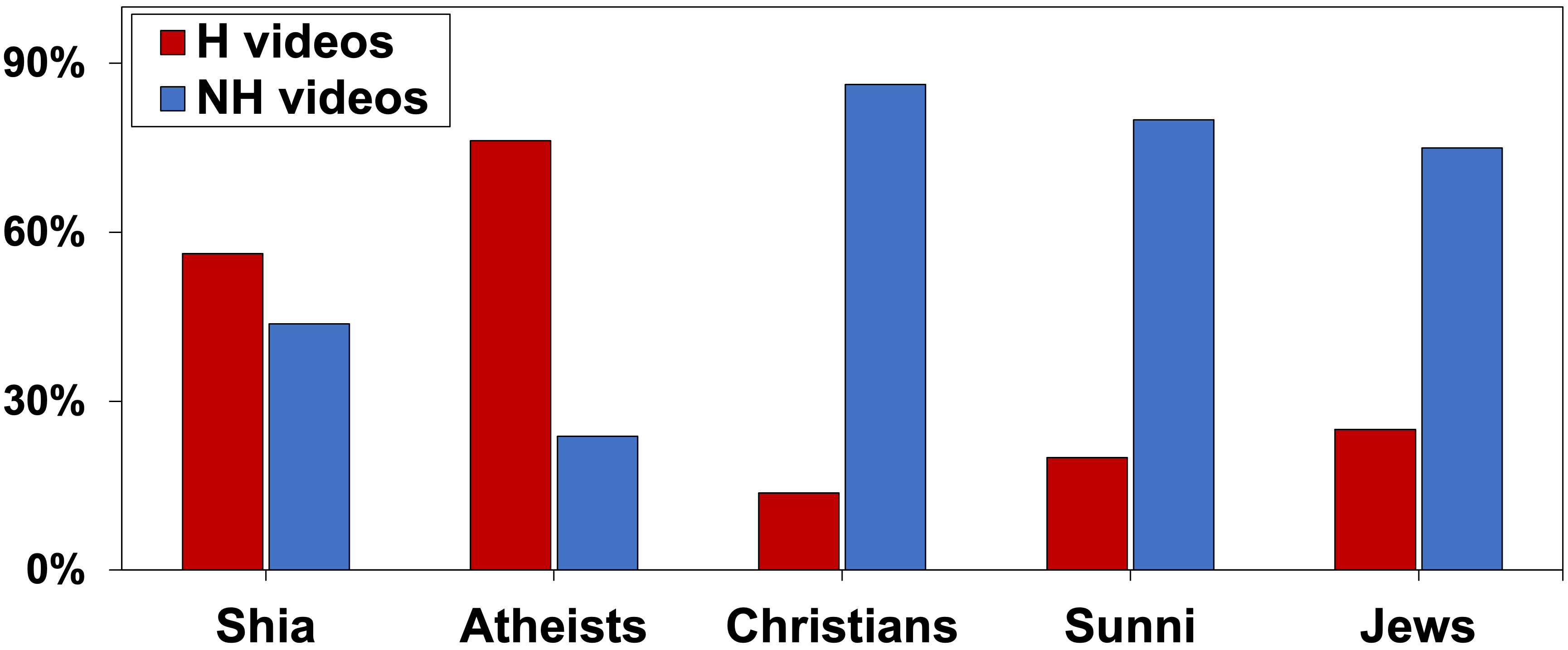}
\label{search_keyword_dist}}

\caption{Distribution of hateful and non-hateful videos by keyword in (a) recommendation audit and (b) search audit.}
\label{search_recom_keyword_dist}
\end{figure}

\textbf{Across Keyword Differences.} We investigated whether the amount of hateful content differed across the five keywords. For recommendation audit videos (Figure \ref{recom_keyword_dist}), we found that the amount of hateful content differed significantly across keywords, ${\chi}^2$(4,4000)=209, $p$<0.001. To find exactly which keyword significantly differed from the other, we followed up with post-hoc pairwise comparisons and used the Benjamini-Hochberg procedure to control FDR at level 0.05. We found that all keywords differed significantly from each other except for the pairs (atheists and Shia) and (Christians and Sunni). As is evident from the graph, watching videos related to Shia resulted in having the highest proportion of hateful video recommendations (43.4\%), followed closely by  atheists (41.8\%). Videos related to Christians and Sunni resulted in a similar proportion (26\%) of hateful video recommendations. Jews-related videos yielded the lowest amount of hateful recommended videos (15.9\%).

For search audit videos (Figure \ref{search_keyword_dist}), we also found a significant difference in the distribution of hateful and non-hateful videos across keywords, ${\chi}^2$(4,400)=97, $p$<0.001. Post-hoc tests with BH-FDR adjustment revealed that the keyword `atheists' returned a significantly higher proportion of hateful videos (76\%) than any other keyword. About 56\% of `Shia' keyword search results were hateful, which was significantly different from all other keyword search results. The proportion of hateful videos didn't significantly differ across the keywords Jews (25\%), Sunni (20\%), and Christians (14\%). \textbf{Insight:} {\it Searching YouTube for Shia and atheists videos result in high volume of hateful videos appearing in YouTube's search results and recommendations.}

\section{Discussion}
%
Our analysis provides proxy indicators of the prevalence of Arabic religiously intolerant videos on YouTube. In our non-personalized analysis, we found that 30\% of videos returned in search results were hateful. This is particularly alarming given that the data collection process was based on entirely innocuous terms. However, we note that the percentage of hateful content found within Arabic religious discussions on Twitter using a similar methodology was considerably larger (42\%) \cite{albadi2018they}. This could be attributed to the fact that creating content on Twitter is much easier than creating content on YouTube. It could also reflect differences in the degree of moderation and hateful content policing the two platforms are conducting. When comparing targets of hate on YouTube versus Twitter, we found that the main difference lies in the extent of hate targeting the Jewish community; while the Jews were the most targeted religious community on Twitter \cite{albadi2018they}, they were among the least targeted religious groups on YouTube. On the other hand, Shia and Atheists remain among the top of the most targeted religious groups on both platforms. 

Our recommendation graph analysis based on nearly 1M captured recommendation transitions between videos suggests that YouTube recommendation algorithm can expose users arriving at non-hateful videos to hateful ones. We also found that 31\% of videos recommended to hateful videos were also hateful, which could reinforce users exposure to radical content. However, these findings don't necessarily support the claim that YouTube recommendation definitely have a role in radicalization as radicalization can occur offline or in other social spaces. A recent research \cite{hosseinmardi2021examining} found that users who consumed far-right radical videos arrived at such videos more frequently from YouTube search results or an external website rather than following YouTube recommendation chains. Additionally, our study takes into account all top four recommended videos, while in a realistic setting users make a subconscious decision on which video to watch next based on  multiple factors (e.g., thumbnail, video title, appearing in Up Next, etc.). Nevertheless, it is fair to conclude that YouTube may have a role in online radicalization by exposing users to radical content through surfacing it in its search results and recommendations. 



Disturbingly, our personalized audits suggest that personalization in general results in surfacing more hateful content and that religious identity and gender could contribute to greater levels of exposure to hateful content. Particularly relevant to our audit experiments is an audit study by \citet{hussein2020measuring}, in which they investigated the effect of watch history and demographics (i.e., age, gender, and geographical location) on the amount of misinformative content returned in YouTube search results and recommendations. They found that in most cases, men were recommended significantly more misinformative videos than women. In line with their findings, we found that male Sunni profiles were recommended significantly more hateful videos than female Sunni profiles. This is concerning, especially in light of the statistics that show males being more prone to radicalization than females \cite{moller2018terrorism}.

In previous audit studies \cite{hussein2020measuring,Kostantinos2022justAflue}, the process of developing a watch history for profiles was conducted automatically using Selenium bots (i.e., automated web browsing bots), in which each profile was controlled by a bot that watched sequentially in one session a predetermined set of videos. However, in our personalized audit experiments, the process of building the watch history was carried out manually in a controlled environment by the first two authors over nine weeks. We believe that our methodology leads to building a more realistic, human-like watch history than the one using bots. On the other hand, when assigning annotations to videos that the profiles get exposed to, we argue that doing so automatically using machine-learning classifiers rather than relying solely on human annotators would facilitate auditing recommendation algorithms at a larger scale.

In 2017, YouTube along with other major social media platforms formed the Global Internet Forum to Counter Terrorism in an effort to make their online services free from hateful and extreme content \cite{YouTube2017}. To assess YouTube's current countermeasures in removing Arabic hateful content, we checked the current status of all videos analyzed in our study after two years from data collection. We found that only 16.44\% of the hateful videos were removed from the platform in these two years. In addition, 13.14\% of non-hateful videos were also removed. While this difference in distribution was found to be significant (${\chi}^2$(1,351262)=348, $p$<0.001), it is important to note that 83.56\% of hateful videos were still available on the platform as of September 2021. Thus, our study signals a need for a more active and effective technique to guard against Arabic hateful content on YouTube. To our knowledge, our study represents a first effort in characterizing, identifying and understanding the spread of Arabic hateful content on YouTube, and we hope that it would serve as a starting point for other researchers to invest more efforts in ultimately making social networks' Arabic content safer and free from radical content.

Although we showed in Section \ref{classifier_results} that our classifier delivered a comparable performance to other YouTube models developed to tackle other related issues, we acknowledge that our classifier performance is still lower than desired. This reflects the inherent difficulty of capturing hate speech as it can be highly contextual. To boost performance it is essential, though expensive, to acquire more accurately labeled data. It is also worth experimenting with other text embedding techniques such as contextualized token embeddings using Bidirectional Encoder Representations from Transformers (BERT). Another limitation is that our dataset for the non-personalized analysis may not be representative of the entire Arabic religious videos on YouTube given that the volume of such content is unknown. Finally, our personalized audit study modeled a limited number of users focusing on those at opposing ends of the religious spectrum. Additionally, in a real setting, a person may be watching other types of videos, not only religious ones, which could impact the recommendation behavior of the system. Thus, auditing YouTube's recommendations by considering a more comprehensive range of users and online behaviors would be a clear direction for future research.



\section{Conclusion}
In this study, we explored the spread of Arabic YouTube videos targeting religious minorities, how often YouTube's recommendations suggest such videos in a general sense, and the extent of biases exacerbated by YouTube's recommendations based on personal attributes. We found that hateful videos are particularly prevalent in search results and first-level recommendations, mainly targeting Shia an atheists. Recommendations to hateful videos comprised about 15\% of overall recommendations. Our personalized audit experiments revealed that gender, religious ideology, and Islamic denomination can contribute to greater exposure to hateful content.

We showed that YouTube's current countermeasures are ineffective in combating Arabic radical content, with about 84\% of identified hateful videos going undetected as of September 2021. YouTube has invested some efforts into tackling harmful content in the English language. For example, YouTube now recommends debunking videos to those watching videos promoting anti-vaccination myths \cite{hussein2020measuring}. A similar effort should also be invested in tackling Arabic radical content, which not only fuels civil wars in the Arab region, but also contributes to terrorism worldwide. Viewers of Arabic content can also have a role in bringing down extreme content by actively reporting it to YouTube and providing counter-narratives in the comment section whenever possible. The findings and resources presented in this paper facilitate enforcing YouTube's hateful content policy, motivate future research in the area, and raise awareness among the public and concerned agencies.


\bibliographystyle{ACM-Reference-Format}
\bibliography{sample-base}

\appendix

\section{Test Questions Used to assess annotators}
\label{appendix}
Content warning: some readers may find hateful videos identified here to be upsetting or disturbing. 
\begin{table}[h]
    \centering
    \resizebox{\columnwidth}{!}{%
    \begin{tabular}{lll|llp{0.22\linewidth}|lll}
    \toprule
        \textbf{video id} & \textbf{hate} & \textbf{targeted religions} & \textbf{video id} & \textbf{hate} & \textbf{targeted religions} & \textbf{video id} & \textbf{hate} & \textbf{targeted religions} \\ 
        \midrule
        yhckohsG9F8 & no &   & \_lCFs\_-4MhA & no &   & df9z6SReHt8 & yes &  Islam \\ 
        8qtoC9l6Q44 & no &   & HNvY65qyCKY & unrelated &   & WSIWbFdMGZI & yes &  Judaism, Christianity, Atheism \\ 
        QFGR-axebT8 & no &   & KhUcmXUuXQA & no &   & Uw4BazBdtZ4 & no &   \\ 
        dVyNS7PzUYM & yes &  Atheism & Vklc-RbEXYk & no &   & bHCtdSy-Nbw & no &   \\ 
        BzxUmfY2zXg & no &   & 3HbaA9kMkEk & no &   & mt8rYzx8u\_0 & no &   \\ 
        LbAeyeGeprg & unrelated &   & VFHd2R-gpQs & no &   & 1Ja6kmTaIk8 & no &   \\ 
        x4jn1lEmIIw & no &   & gHN1elpukjc & unrelated &   & Q5eJe9KOokc & no &   \\ 
        h\_a3zDMW\_P0 & no &   & mT38cbTw1Cg & no &   & NLHMxeBE9Gg & no &   \\ 
        Q0uLPDR9MMs & yes &  Judaism, Shia & BfXjTFRWjDU & no &   & xpmIAR\_cIlA & no &   \\ 
        hRMEY40HYzE & unrelated &   & 99vGew\_NWaM & no &   & JZwgqYntWtc & unrelated &   \\ 
        tkefR-Nc40A & yes &  Shia & CuPStQ-j7bg & no &   & 9\_ARBi7KuRI & unrelated &   \\ 
        FVcdVREEXL4 & yes &  Shia & jHyOWcbx7C0 & no &   & fNPPeSAk06Q & unrelated &   \\ 
        io1-szAAgz8 & unrelated &   & Ejv30YVRo6c & no &   & hSMWbX\_sPmM & unrelated &   \\ 
        fRSoYn3iEcQ & yes &  Shia & o8E0JB-IoQc & no &   & 0t5G4NWAtgA & unrelated &   \\ 
        2mqL-7kaKKk & no &   & 5AkAGc5nOXw & unrelated &   & DyUb2ixHTFU & unrelated &   \\ 
        DWo4wGuIq\_Y & unrelated &   & e-G7Tt3-P1Q & unrelated &   & b0lh4VYzdL4 & unrelated &   \\ 
        fpnmULVBnUU & no &   & QwMTpa3IcFc & no &   & 82B38Npy38s & unrelated &   \\ 
        AxtJAK6f5Mk & no &   & gkMfpXR\_aro & no &   & myUAndPEqCA & no &   \\ 
        d8LgcHyhyp4 & unrelated &   & o\_A4-TI5jFY & unrelated &   & 7fDMSoJghhM & yes &  Shia \\ 
        gMK12E37FxE & yes &  Shia & Iw0cNWeHoTE & no &   & ZQ\_iqO-dcpg & no &   \\ 
        FUx\_IyVGUAE & unrelated &   & GCtt21GPaUc & unrelated &   & ISuvq\_MdeT4 & unrelated &   \\ 
        xbpp7eDqkSk & yes &  Judaism & spiiOmbDDmA & no &   & yzkLgpe5BQw & no &   \\ 
        tzXx62NEhbY & no &   & GGrmb6UTleo & unrelated &   & S93XXoXgwog & no &   \\ 
        hO7Wer85jyg & no &   & r\_bdGmJopkA & unrelated &   & gLY3fw0fsSo & yes &  Shia \\ 
        XTLla\_Mc6EU & yes &  Atheism & EBALA4xESSM & unrelated &   & LVjD28UAjdE & no &   \\ 
        JtAMz7bmXXg & yes &  Atheism & MF86I\_Cj71c & no &   & DlQojAGZu\_8 & no &   \\ 
        EjhIwOBcd88 & yes &  Atheism & Bnxoox6HB5w & yes &  Judaism, Christianity & HM1V\_9U6h6M & no &   \\ 
        Qm3ghn7tgjA & yes &  Shia & zJ7dFGAHvSc & yes &  Christianity & ZyLvudOqDG0 & yes &  Christianity \\ 
        308v3Rw-3Vo & yes &  Shia & S7nvJY3ng2w & yes &  Shia & bY3Jvb6-JHw & yes &  Atheism \\ 
        5rYQ-cUAe9I & no &   & ZYpBzSgUFNY & unrelated &   & gukyE6PD9ec & yes &  Judaism, Islam \\ 
        MKzzrth6T\_8 & no &   & EBN1OpRwM-c & no &   & AYN356BEbfw & no &   \\ 
        HsjhJz4mpVs & yes &  Shia & DOeVkBMkl\_0 & unrelated &   & sEEOoSuMVvs & no &   \\ 
        nAP8wMCCe5k & yes &  Shia & BpuQvOGj5qM & yes &  Shia & CvYcnpH4Gyg & unrelated &   \\ 
        FKbg2Lh0OFg & yes &  Shia & KkFeB1vYh\_k & yes &  Atheism & 51NJ577mVHA & no &   \\ 
        Sn1KOSgLTHM & no &   & EoekkBYobYs & no &   & bTCr6VOUkJk & yes &  Judaism, Christianity, Islam \\ 
        GffdGf1YPSY & no &   & 7ShmzK-35ZQ & no &   & bX8Tae3cB-8 & yes &  Atheism \\ 
        FREy-imMDK8 & unrelated &   & r-IVThm8InE & no &   & OWOB0tDfY7M & yes &  Judaism, Christianity \\ 
        mCXvytpTYyQ & no &   & SwH5QQn34Ws & unrelated &   & ndIugDoFOlo & yes &  Shia, Sunni \\ 
        9GXC5r\_SSgg & unrelated &   & ZfL3AHF4DYA & unrelated &   & s3ISnNsNX3A & yes &  Shia \\ 
        gjM3cFybQFw & unrelated &   & xc-9xVGICkk & unrelated &   & t-u-M5jD-84 & yes &  Judaism \\ 
        h\_ujLmijvEI & unrelated &   & qtNwwiavyaY & yes &  Atheism & 6gSM1GUpcng & yes &  Shia \\ 
        F-AIYsYDjXQ & unrelated &   & TGv7AEF9K38 & yes &  Shia & UhrpQeWhdNk & unrelated &   \\ 
        yXCxNhwYKUQ & unrelated &   & F2DiJP1kgdU & yes &  Shia & guZ9RqcEdjw & unrelated &   \\ 
        3W45TxFAas4 & unrelated &   & Rwq4rD\_qH\_4 & yes &  Shia & W3\_Et2yzOTE & unrelated &   \\ 
        Z\_-n4E9pkmY & unrelated &   & w6akRu0VEm8 & yes &  Shia & Z7ldz6Ytw40 & unrelated &   \\ 
        rXs6iJYfulM & no &   & g2iS61WCsvE & yes &  Shia & LKOoPcHxPhU & unrelated &   \\ 
        8Ax1tOZxf7g & no &   & V7p0e7gJn4M & yes &  Shia & gHoY2AU4snc & unrelated &   \\ 
        E6sNbYsJ6R8 & no &   & 1m4K4f6YYE0 & no &   & sQilyC\_ICyU & unrelated &   \\ 
        JlJeOeNSjlU & no &   & piJ8ro331Ak & no &   & w8Digjy8QbI & unrelated &   \\ 
        PLRapMO-VUQ & yes &  Shia, Sunni & sKwAPfnUwb0 & no &   & 09P4NCJs3uU & unrelated &   \\ 
        16ntPL7Ij\_0 & unrelated &   & bSR\_G-Ic\_Jk & no &   & YQ\_3OAzWjQk & unrelated &   \\ 
        u6b77OcRVUs & yes &  Shia & 5C-Rf61hU5o & unrelated &   & J5aseBw4BmM & unrelated &   \\ 
        2a4eIM7xIvY & no &   & n5Sz8XlOpmY & unrelated &   & cr6bh3xUWMg & unrelated &   \\ 
        mvUw6Vw1NeA & yes &  Christianity & xlutMLEp40Y & unrelated &   & hHBdYzpRW48 & yes &  Shia \\ 
        kohMe2iew\_4 & no &   & GtEal9rT3RM & yes &  Judaism, Christianity & X5IkWjg1huA & yes &  Atheism \\ 
        TSj2mLnqR-c & no &   & KWz8ON4vvds & unrelated &   & 2IGnJVY40vw & yes &  Christianity \\ 
        tFZ\_rwHIyO0 & no &   & AG8D\_wiH7KE & unrelated &   & DW2Zkz\_zl-o & unrelated &   \\ 
        UNWCrxr3Tfc & no &   & KbWD0X9bqrA & unrelated &   & qQOLPNyssGE & unrelated &   \\ 
        Gxst3LPF86U & no &   & 3JsB6cuVZu0 & unrelated &   & xPSv-ThlK\_4 & unrelated &   \\ 
        NkvldtFqqNs & no &   & CSM0HR0aFWs & unrelated &   & TYdyg4XvRQ0 & no &   \\ 
        9gVOOW9HH1c & no &   & YjtitQtJzNk & no &   & ZekNtF8RV-A & unrelated & \\
        g7O7cOmOLTE & no &  & 1dNDGVHHm2Y & yes &  Judaism, Christianity, Atheism, and others  & & &  \\ 
   \bottomrule
    \end{tabular}
    }
\end{table}

\end{document}